\documentclass[aps,superscriptaddress,notitlepage,onecolumn,bibnotes,
]{revtex4-1}
\usepackage[utf8]{inputenc}
\usepackage{graphicx}
\usepackage{float}
\usepackage{gensymb}
\usepackage{upgreek}
\usepackage{color,amsmath}
\usepackage[colorlinks,linkcolor=blue,citecolor=blue,urlcolor=blue]{hyperref}
\usepackage[normalem]{ulem}
\usepackage{siunitx}
\usepackage{enumitem}
\usepackage{ulem}
\usepackage{amsmath,amsfonts}
\usepackage{lineno}
\usepackage{ragged2e}
\usepackage{textcomp}
\usepackage{color,soul}
\usepackage{textgreek}
\usepackage{hyphenat}
\usepackage{adjustbox}
\usepackage{setspace}
\usepackage{fancyhdr}
\usepackage{natbib}
\setcitestyle{numbers}  
\usepackage{natbib}
\usepackage[table,xcdraw]{xcolor}

\renewcommand{\cite}[1]{\textsuperscript{\citenum{#1}}}

\renewcommand{\doi}[1]{}
\renewcommand{\url}[1]{}

\makeatletter
\renewcommand\section{\@startsection {section}{1}{\z@}%
   {-2.0ex \@plus -1ex \@minus -.2ex}%
   {1.0ex \@plus.2ex}%
   {\normalfont\bfseries}}
\renewcommand\subsection{\@startsection {subsection}{2}{\z@}%
   {-1.5ex \@plus -1ex \@minus -.2ex}%
   {0.5ex \@plus .2ex}%
   {\normalfont\bfseries}}
\makeatother

\DeclareUnicodeCharacter{2212}{-}

\newcommand{\fmarki}{*}
\def\@fnsymbol#1{{\ifcase#1\or \fmarki  \else\@ctrerr\fi}}
\makeatother

\begin{document}

\setlength{\parskip}{0.55em}  

\title{\large Polarization boost and ferroelectricity down to one unit cell \\ in layered Carpy-Galy La\textsubscript{2}Ti\textsubscript{2}O\textsubscript{7} thin films}
\author{Elzbieta Gradauskaite\footnote{Corresponding author: \href{mailto:elzbieta.gradauskaite@cnrs-thales.fr}{elzbieta.gradauskaite@cnrs-thales.fr}}}
\affiliation{Laboratoire Albert Fert, CNRS, Thales, Universit\'e Paris Saclay, 91767 Palaiseau, France}
\author{Anouk S. Goossens}
\affiliation{Laboratoire Albert Fert, CNRS, Thales, Universit\'e Paris Saclay, 91767 Palaiseau, France}
\author{Xiaoyan Li}
\affiliation{Laboratoire de Physique des Solides, CNRS, Universit\'e Paris Saclay, 91405 Orsay, France}
\author{\hbox{Lucía Iglesias}}
\affiliation{Laboratoire Albert Fert, CNRS, Thales, Universit\'e Paris Saclay, 91767 Palaiseau, France}
\author{Alexandre Gloter}
\affiliation{Laboratoire de Physique des Solides, CNRS, Universit\'e Paris Saclay, 91405 Orsay, France}
\author{Quintin N. Meier}
\affiliation{Université Grenoble Alpes, CNRS, Institut Néel, 38042 Grenoble, France}
\author{Manuel Bibes}
\affiliation{Laboratoire Albert Fert, CNRS, Thales, Universit\'e Paris Saclay, 91767 Palaiseau, France}

\begin{abstract}
Layered perovskite-based compounds offer a range of unconventional properties enabled by their naturally anisotropic structure. While most renowned for the superconductivity observed in the Ruddlesden-Popper phases, many of these layered compounds are also ferroelectric and exhibit a sizeable in-plane polarization. Among these, the Carpy-Galy phases (\textit{A}\textsubscript{\textit{n}}\textit{B}\textsubscript{\textit{n}}O\textsubscript{3\textit{n}+2}) characterized by 110-oriented perovskite planes interleaved with additional oxygen layers have been debated as platforms for hosting not only a robust polarization but also multiferroicity and polar metallicity. However, the challenges associated with the synthesis of ultrathin Carpy-Galy films and understanding the impact of strain on their properties limit their integration into devices.  Addressing this issue, our study focuses on La\textsubscript{2}Ti\textsubscript{2}O\textsubscript{7}, an \textit{n}=4 (\textit{A}\textsubscript{2}\textit{B}\textsubscript{2}O\textsubscript{7}) representative of the Carpy-Galy family, exploring its growth and concurrent phase stability on various substrates under different strain conditions. Remarkably, we demonstrate that a 3\% tensile strain from DyScO\textsubscript{3} (100) substrates promotes a controlled layer-by-layer growth mode, while SrTiO\textsubscript{3} (110) and LaAlO\textsubscript{3}-Sr\textsubscript{2}TaAlO\textsubscript{6} (110) that exert negligible and compressive strains, respectively, require post-deposition annealing to achieve similar results. Using scanning probe microscopy, X-ray diffraction, scanning transmission electron microscopy, and polarization switching experiments, we confirm that these films possess exceptional ferroelectric properties, including a polarization of 18 \textmu C/cm\textsuperscript{2} — more than three times higher than previously reported — as well as persistence of ferroelectricity down to a single-unit-cell thickness. This study not only advances our understanding of Carpy-Galy phases in thin films but also lays a foundation for their application in advanced ferroelectric device architectures.
\end{abstract}
\maketitle

\section{Introduction} 
To date, extensive research on low-energy, non-volatile ferroelectric memory has predominantly focused on classical perovskite oxides. The polar behavior in these materials arises from the displacement of cations with respect to anions, leading to an off-centering of positive charges within their oxygen octahedral cages. The straightforward and nearly cubic structure of perovskite oxides simplifies their fabrication into epitaxial thin films, facilitating their integration into complex epitaxial heterostructures. However, perovskite oxides are prone to finite-size effects\cite{Junquera2003} and exhibit no pronounced anisotropy in their properties, which restrict their scalability in nanodevice applications. These limitations have driven the search for alternative materials with more durable polar properties that can combine additional functionalities within the same phase. Among non-traditional ferroelectrics, such as improper ferroelectrics\cite{li_hexagonal_2020, Bousquet2008}, HfO\textsubscript{2}-based ferroelectric thin films\cite{Park2018}, two-dimensional polar materials\cite{Guan2020}, organic-inorganic hybrid ferroelectrics\cite{xu_hybrid_2019}, layered ferroelectrics\cite{benedek_understanding_2015} are distinguished by their high structural flexibility and robust ferroelectric properties.

Layered ferroelectrics\cite{benedek_understanding_2015} consist of periodically repeating perovskite-like slabs interleaved with spacers that carry localized ionic charges, significantly influencing the electrostatic boundary conditions. Ferroelectricity in these structures is primarily stabilized by the confinement effects of their large periodic unit cells rather than by electronic hybridization within the \textit{AB}O\textsubscript{3} units, thereby enhancing their stability and robustness compared to traditional ferroelectric perovskites. Historically, layered ferroelectrics have been challenging to synthesize, limiting their study mostly to bulk crystals or ceramics. Recent advances, however, have facilitated the preparation of these materials in thin-film form through molecular beam epitaxy and pulsed laser deposition (PLD). It has been demonstrated that epitaxial stabilization of layered ferroelectrics can lead to unprecedented functionality, such as lack of critical thickness for ferroelectricity\cite{Keeney2020a,Gradauskaite2020a,gradauskaite2021a, Keeney2020}, charged domain walls\cite{Gradauskaite2020a,Gradauskaite2022a,Moore2022}, polar vortex topologies\cite{Moore2022}, and significantly improved robustness against ferroelectric fatigue\cite{Gradauskaite2020a}. Additionally, integrating layered ferroelectrics into canonical perovskite heterostructures has emerged as a novel approach to mitigate depolarizing field effects and enable the formation of unique polar structures characterized by their chirality\cite{gradauskaite_defeating_2023}.

The most extensively studied families of layered ferroelectrics as epitaxial films are the Ruddlesden-Popper phases, due to their superconducting properties\cite{bednorz_possible_1986, maeno_superconductivity_1994}, and the Aurivillius phases, known for their exceptional resistance to ferroelectric fatigue\cite{a-paz_de_araujo_fatigue-free_1995}. However, the layered \textit{A}\textsubscript{\textit{n}}\textit{B}\textsubscript{\textit{n}}O\textsubscript{3\textit{n}+2} ferroelectrics\cite{carpy_systeme_1974, nanamatsu_crystallographic_1975,isupovCrystalChemicalAspects1999,lichtenbergSynthesisPerovskiterelatedLayered2001,lichtenbergSynthesisStructuralMagnetic2008}, recently termed the Carpy Galy (CG) phases\cite{nunez_valdez_origin_2019}, have been rarely investigated as thin films. Characterized by 110-oriented perovskite planes interleaved with additional oxygen layers, the CG phases exhibit unparalleled ferroelectric Curie temperature (T\textsubscript{c}) values of up to 2000\textdegree{}C -- particularly for \textit{n}=4 ferroelectrics -- and substantial polarization. The dominant ferroelectric instability in the CG phases originates from oxygen octahedra rotations that are prevalent in many magnetically ordered non-polar perovskites. The truncation of the unit cell, however, implies that such oxygen octahedra rotations sum up to a non-zero net polarization in the CG unit cell, a phenomenon coined as proper topological ferroelectricity\cite{lopez-perezInitioStudyProper2011}. Additionally, CG compounds have been debated to be potential material platforms for hosting incommensurate phases\cite{daniels_incommensurate_2002,howiesonIncommensurateCommensurateTransition2020}, multiferroicity\cite{lopez-perezInitioStudyProper2011, ederer_electric-field-switchable_2006}, as well as polar metallicity\cite{Kuntscher2004}. Despite their considerable potential, synthesizing high-quality single-crystalline CG thin films poses significant challenges and a thorough understanding of their ferroelectric properties under diverse epitaxial strain states is still lacking, hindering the optimization of these materials for practical applications.

Here we report on the synthesis of uniaxial in-plane-polarized single-crystalline La\textsubscript{2}Ti\textsubscript{2}O\textsubscript{7} (LTO) \textit{n}=4 thin films of the Carpy-Galy (CG) phase on various substrates that exert negligible, tensile, and compressive epitaxial strains. In-situ monitoring with reflection high-energy electron diffraction (RHEED) allows us to follow the growth dynamics and uncover how epitaxial strain influences the layering of the CG phase in thin films. Remarkably, we observe that while negligible epitaxial strain results in the layer-by-layer coverage of pseudo-perovskite blocks with disordered oxygen planes, applying some epitaxial strain is crucial for stabilizing a true layer-by-layer growth mode, during which full \textit{A}\textsubscript{2}\textit{B}\textsubscript{2}O\textsubscript{7} electroneutral units coalesce to form a continuous film. Notably, substantial tensile strain within the film plane facilitates phase-pure, layer-by-layer growth for films up to tens of nanometers thick without the need for post-deposition annealing. This observation is rationalized by the relationship between polar distortion modes and different epitaxial strains through density-functional theory (DFT) calculations. The combination of X-ray diffraction (XRD), scanning transmission electron microscopy (STEM), piezoresponse force microscopy (PFM), and ferroelectric switching experiments allows us to confirm the synthesis of epitaxial single-crystalline LTO films. The importance of epitaxial strain in stabilizing high-quality thin films is exemplified by more than a three-fold enhancement in ferroelectric polarization compared to the literature values reported for bulk single crystals. The films exhibit two-variant nanoscale ferroelectric domain patterns characterized by clearly defined uniaxial in-plane polarization. Lastly, we demonstrate the absence of critical thickness for ferroelectricity in the films of the CG phase and reveal the domain formation mechanism related to the out-of-phase boundary formation at the substrate interface. These insights pave the way for controlled synthesis of phase-pure ferroelectric CG thin films, opening avenues for their integration into functional perovskite-based heterostructures.

\section{Results and Discussion} 

\subsection{In-situ monitoring of strain-driven layering in epitaxial LTO films}

\begin{figure}[b!]
  \centering 
  \includegraphics[width=0.7\textwidth, clip, trim=4 4 4 5]{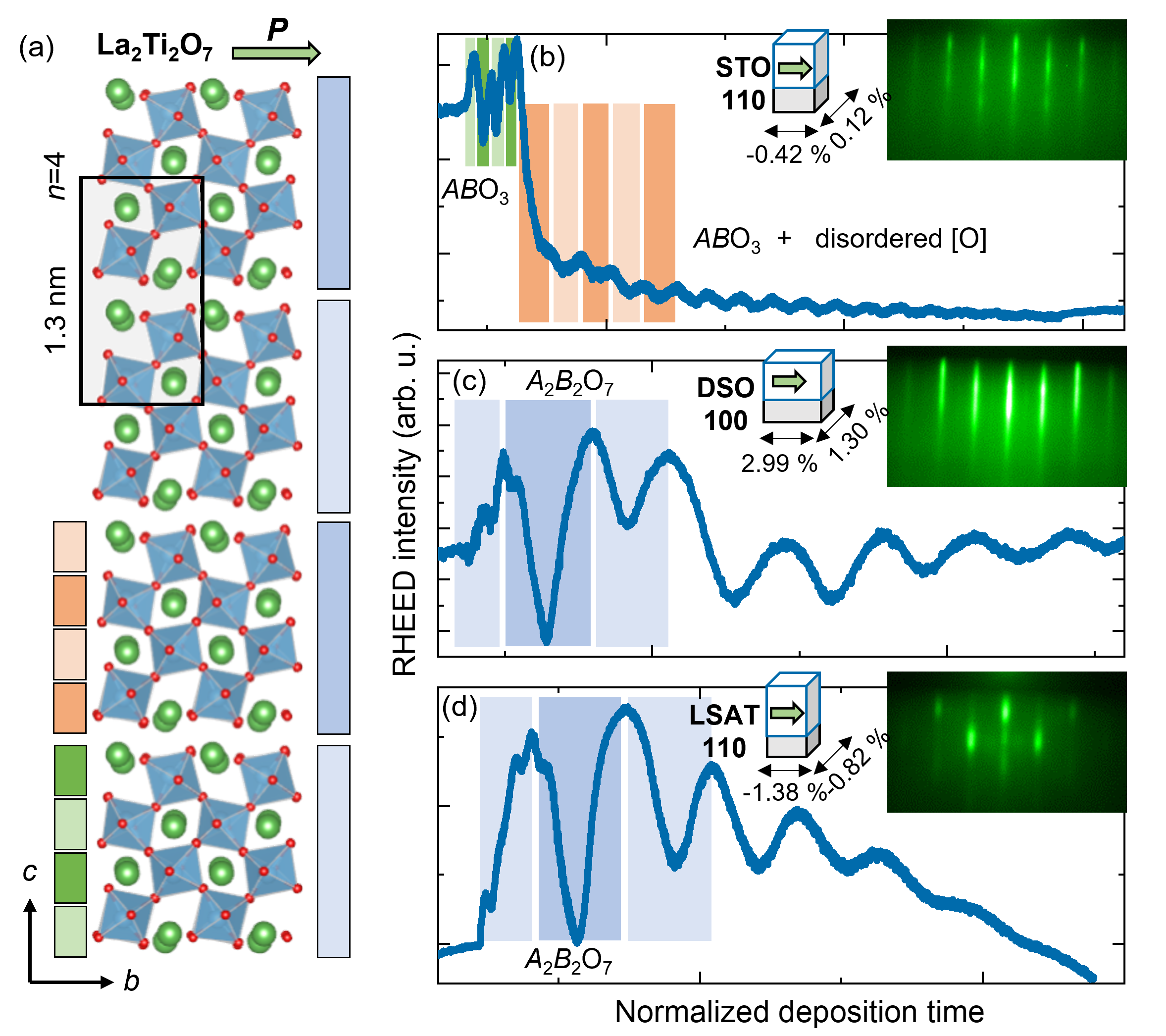}
  \caption{\textbf{RHEED monitoring of epitaxial LTO film growth.} Schematic of the LTO crystal structure viewed along the \textit{b-c} plane, showing uniaxial polarization along the \textit{b}-axis. The rectangular box represents one unit cell, which has a height of 1.3 nm along the out-of-plane \textit{c}-axis. (b-d) RHEED intensity traces for LTO films grown on three substrates at 800\textdegree{}C, partial oxygen pressure of $6\cdot10^{-2}$ mbar, laser repetition rate of 1 Hz with fluence of 1.2 J cm\textsuperscript{-2}: (b) STO (110) with negligible strain, (c) DSO (100) with tensile strain, and (d) LSAT (110) with compressive strain. Insets display the corresponding epitaxial strain orientations along the in-plane axes and the observed RHEED patterns for each substrate, confirming two-dimensional growth.}\label{fig:rheed}
\end{figure}

For our study, we selected LTO as a representative \textit{n}=4 ferroelectric\cite{nanamatsuNewFerroelectricLa2Ti2o71974}  of the CG phase, with the empirical formula \textit{A}\textsubscript{2}\textit{B}\textsubscript{2}O\textsubscript{7}. LTO, likely the most investigated ferroelectric in the CG family, possesses a monoclinic unit cell (space group $P2_1$)\cite{schmalle_twin_1993} and a ferroelectric Curie temperature (T\textsubscript{c}) of about 1500\textdegree{}C\cite{nanamatsuNewFerroelectricLa2Ti2o71974}. The crystallographic \textit{c}-axis is oriented perpendicular to the oxygen spacers, the \textit{a}-axis aligns along infinite chains of oxygen octahedra, and the \textit{b}-axis extends through the opposite edges of these octahedra\cite{isupovCrystalChemicalAspects1999} (refer to Figure 1a). Notably, LTO exhibits a net polarization along the \textit{b}-axis\cite{nanamatsuNewFerroelectricLa2Ti2o71974, nunez_valdez_origin_2019}. To fabricate thin films with in-plane polarization, we aim to stabilize single-crystalline 001-oriented LTO on three substrates: SrTiO\textsubscript{3} (STO) (110), DyScO\textsubscript{3} (DSO) (100), and LaAlO\textsubscript{3}-Sr\textsubscript{2}TaAlO\textsubscript{6} (LSAT) (110), which impart negligible, tensile, and compressive in-plane epitaxial strain to the LTO unit cell, respectively (see Supporting Note S1 and insets in Figure \ref{fig:rheed}b-d).

In order to optimize PLD growth conditions of epitaxial LTO thin films we used RHEED in-situ monitoring, which allowed us to track the film roughness and its crystallinity with respect to the deposition time, and to determine whether the growth is two- or three-dimensional. For STO (110) substrates, which impart minimal strain, achieving high-quality films required high substrate temperature, low oxygen partial pressure, and a low laser repetition rate, consistent with earlier reports\cite{havelia2009,ohtomoEpitaxialGrowthElectronic2002,haveliaGrowthLa2Ti2o7LaTiO32008,havelia2009a,kasparTuningPiezoelectricProperties2018}. Figure \ref{fig:rheed}b displays the time-dependent intensity of the RHEED signal for an LTO film with oscillations confirming a layer-by-layer growth mode, followed by a clear drop in intensity after the first four oscillations (see caption for deposition parameters). Ex-situ X-ray reflectometry (XRR) reveals that each RHEED oscillation corresponds to the deposition of approximately one perovskite-like \textit{AB}O\textsubscript{3} layer. The higher intensity oscillations at the beginning of the growth, with a consistent count of four, point to the initial growth of LaTiO\textsubscript{3} layers at the substrate interface (marked in green). Once this thickness is exceeded, in accordance with the stoichiometry of the ceramic target, oxygen planes are introduced and seem to increase the surface roughness, explaining the observed drop in RHEED intensity (orange).

Next, we investigated the impact of epitaxial strain on film growth. Remarkably, when LTO film deposition was performed under identical conditions but with tensile in-plane strain provided by the DSO (100) substrate, we observed distinctively different growth dynamics, as depicted in Figure \ref{fig:rheed}c. The RHEED oscillations display higher intensity amplitude and significantly lower frequency (marked in blue), with one oscillation corresponding to every four on the STO (110) substrate (cf. Figure \ref{fig:rheed}b). Post-deposition XRR measurements unambiguously relate each RHEED oscillation to a full \textit{A}\textsubscript{2}\textit{B}\textsubscript{2}O\textsubscript{7} unit cell of the CG phase, rather than to the pseudo-perovskite \textit{AB}O\textsubscript{3} units observed during deposition on STO (110). This indicates the formation of islands comprising the entire electroneutral LTO unit cell, each 1.3 nm in height, which subsequently coalesce to create atomically flat full-layer coverage. Lastly, we repeat the same deposition of the LTO film on the LSAT (110) substrate, inducing compressive in-plane strain. Similarly to the deposition on DSO, we record pronounced coalescent layer-by-layer oscillations with a similar period corresponding to the deposition of one full LTO unit cell (marked in blue), see Figure \ref{fig:rheed}d. However, the rapidly decreasing RHEED intensity and fading oscillations suggest that this growth mode is sustainable only up to 6 unit cells under this strain state, eventually transitioning to a three-dimensional island growth as captured in the RHEED pattern (inset of Figure \ref{fig:rheed}d).

\subsection{Tensile strain as a way to stabilize CG phase amid competing phases in LTO films}

\begin{figure}[t!]
  \centering 
  \begin{adjustbox}{width=1.05\textwidth, center}
    \includegraphics[width=1.05\textwidth]{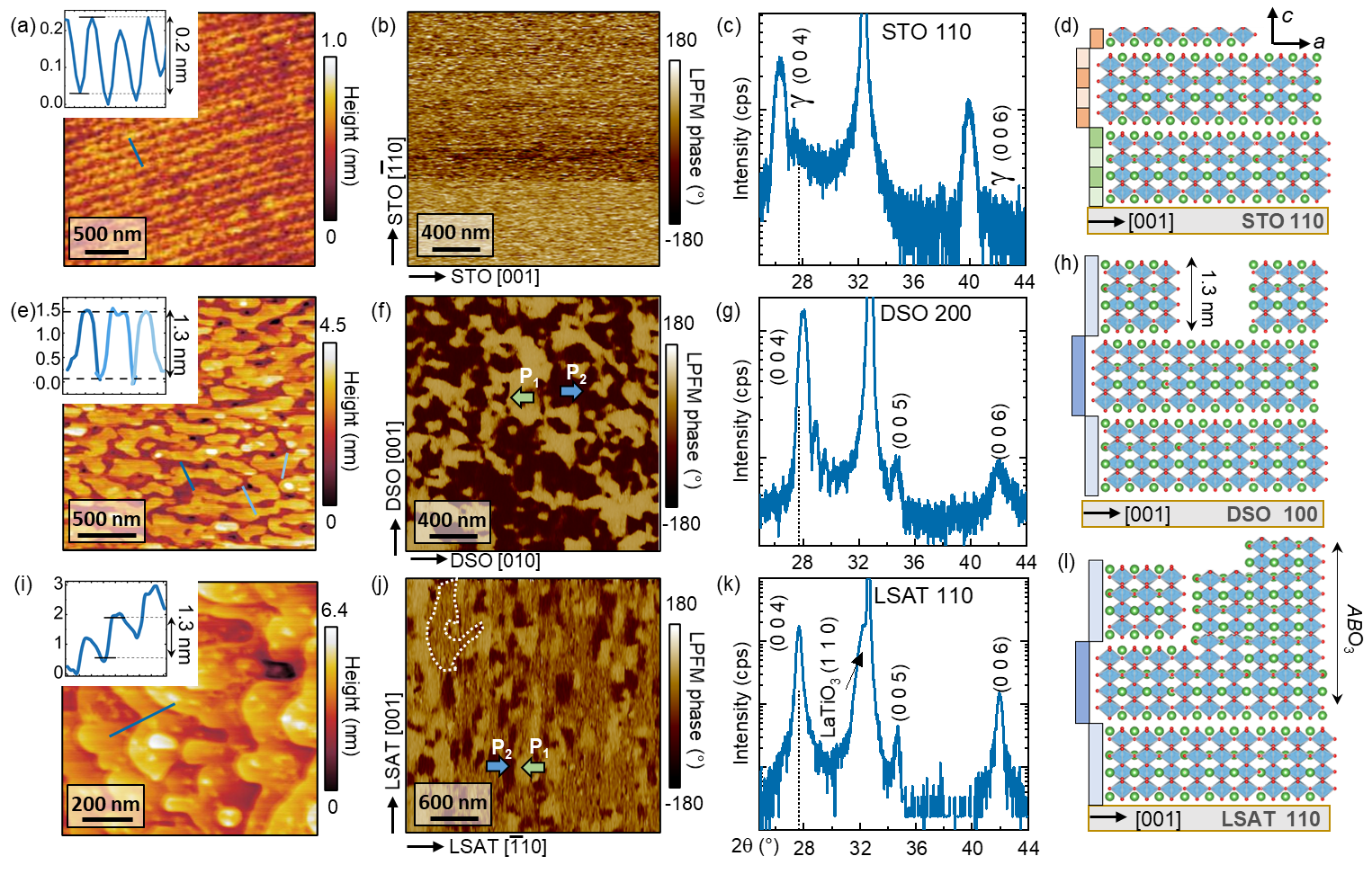}
  \end{adjustbox}
  \caption{\textbf{Elucidating LTO growth modes on STO, DSO, and LSAT substrates through structural and ferroelectric analysis.} (a-d) For LTO films on STO (110): (a) AFM image showing atomically flat surface, (b) PFM image indicating lack of ferroelectric polarization, (c) XRD diffractogram showing the presence of a metastable $\gamma$ phase, and (d) schematic of the corresponding growth mode. (e-h) On DSO (100): (e) AFM image displaying islands of 1.3 nm in height, (f) PFM image revealing in-plane-polarized ferroelectric domains, (g) XRD pattern showing 00\textit{l}-oriented CG phase and Laue fringes indicative of high-quality film, and (h) growth schematic illustrating coalescent layer-by-layer mode. (i-l) On LSAT (110): (i) AFM image of flat step terraces with a height of 1.3 nm, (j) PFM image with three contrast levels suggesting mixed-phase regions, (k) XRD diffractogram identifying the CG and perovskite phases, and (l) corresponding growth schematic.}\label{fig:pfm}
\end{figure}

To shed light on the marked differences in the LTO growth dynamics on the three different substrates, we employ atomic force microscopy (AFM) to evaluate the resulting surface morphology, PFM to measure their local ferroelectric response, and XRD to investigate their macroscopic crystallinity. The films on STO are atomically flat (see Figure \ref{fig:pfm}a), which is consistent with the layering of perovskite units. The lateral PFM (LPFM) measurements (see Figure \ref{fig:pfm}b), however, show no signal indicative of ferroelectric polarization or the existence of ferroelectric domains. This result is further corroborated by XRD showing no 001-oriented peaks corresponding to the CG phase (the theoretical\cite{schmalle_twin_1993} position of the most intense LTO 004 reflection is indicated with a dashed line). The peaks are shifted to lower $2\Theta$ values and can instead be associated with the reflections of orthorhombic, kinetically-favored metastable $\gamma$ polymorph\cite{havelia2009, bayart2013} (see Figure \ref{fig:pfm}c). This suggests that while the coverage with perovskite-like planes was taking place in a controlled layer-by-layer fashion, the oxygen planes were not provided enough energy to crystallize in an ordered manner, i.e.\ every four perovskite layers, see schematic in Figure \ref{fig:pfm}d. This observation goes hand in hand with the work of Havella \textit{et al.}\cite{havelia2009}, which described the competition between the more thermodynamically stable CG phase and the kinetically favored metastable $\gamma$ polymorph. We find that to achieve a layered \textit{A}\textsubscript{2}\textit{B}\textsubscript{2}O\textsubscript{7} phase on STO (110) without oxygen plane disorder, the films must be deposited at very high substrate temperatures (above 1000\textdegree{}C). However, at these temperatures, the adsorbed species gain excess kinetic energy, resulting in a step-flow growth mode. This mode produces no RHEED oscillations, limiting the ability to monitor film thickness with unit-cell precision. Additionally, the high kinetic energy, combined with the splashing of ejected target particles, leads to rougher surfaces, further complicating film quality control, see Supporting Note S2.

The post-deposition AFM measurements of films grown on DSO (Figure \ref{fig:pfm}e) reveal precisely 1.3 nm-high islands on atomically flat surfaces, confirming the coalescence of complete \textit{A}\textsubscript{2}\textit{B}\textsubscript{2}O\textsubscript{7} unit cells. The LPFM measurement (Figure \ref{fig:pfm}f) shows well-defined nanoscale domains with two polarization directions along the \textit{b} polar axis, indicating that the tensile-strained film is stabilized in the ferroelectric phase (see Supporting Note S3 for vector PFM analysis). Additionally, the XRD diffractogram in Figure \ref{fig:pfm}g confirms the pure 001-oriented CG phase, free of any parasitic phases, with Laue thickness fringes around the 004 reflection, highlighting high film quality and sharp interfaces. The coalescent layer-by-layer growth mode, illustrated in Figure \ref{fig:pfm}c, accounts for the high crystallinity of the Carpy-Galy LTO films on DSO. This mode—where unit cells initially grow as isolated islands that later merge into flat layers—mirrors growth behavior observed in Aurivillius phase films\cite{Gradauskaite2020a,gradauskaite2021a}. Remarkably, this growth mode is sustained on DSO even at high substrate temperatures, achieving optimal crystallinity at 1075\textdegree{}C, unlike the lattice-matched STO substrate, where elevated temperatures favor a step-flow growth mode.

At first glance, atomically flat step terraces of 1.3 nm in height (shown in Figure \ref{fig:pfm}i) suggest the successful stabilization of the CG LTO phase on LSAT substrates. Nevertheless, the PFM measurements reveal three contrast levels (Figure \ref{fig:pfm}j): while the brighter and darker areas correspond to the two in-plane-polarized domain variants previously observed for the LTO domains on DSO (cf.\ Figure \ref{fig:pfm}f), the intermediate signal suggests that those areas (e.g.\ the one outlined with a dashed line) are not ferroelectric. This is further supported by XRD results (Figure \ref{fig:pfm}k), which show not only the CG 00\textit{l} reflections but also a significant presence of the 110-oriented non-polar LaTiO\textsubscript{3} perovskite phase in the pristine films. Compressive strain is likely to unfavorably affect the in-plane polarization in LTO, thereby reducing the phase stability compared to the 110-oriented perovskite phase. All this gives insights into the LTO growth mode on LSAT (schematized in Figure \ref{fig:pfm}l): deposition begins with a coalescent layer-by-layer growth mode, but as thickness increases, LaTiO\textsubscript{3} perovskite units begin to incorporate, leading to a suppression of RHEED oscillations and the emergence of three-dimensional growth patterns in RHEED (see Figure \ref{fig:rheed}d). On LSAT this growth behavior is consistent across deposition temperatures ranging from 800\textdegree{}C to 1100\textdegree{}C.

Together, these results from Figures \ref{fig:rheed} and \ref{fig:pfm} indicate that a finite amount of epitaxial strain, either tensile or compressive, is crucial for stabilizing the layer-by-layer growth mode, where the full \textit{A}\textsubscript{2}\textit{B}\textsubscript{2}O\textsubscript{7} layers coalesce into atomically flat layers. This behavior is highly unusual when compared to traditional perovskite ferroelectrics, which typically achieve optimal crystallinity with minimal lattice mismatch. Notably, the CG phase is most effectively stabilized on DSO substrates that exert a 2.99\% tensile strain along the polar \textit{b}-axis. Such significant strain is rarely accommodated well in epitaxial films; however, in this case, it not only supports growth but also enhances the proper layering of the CG phase and its associated ferroelectric polarization.

\subsection{Sequential phase transitions, symmetry-adapted distortions and their interplay with epitaxial strain}\label{sec:DFT}

\begin{figure}[b!]
    \centering
    \includegraphics[width=0.7\linewidth, clip, trim=1 4 4 4]{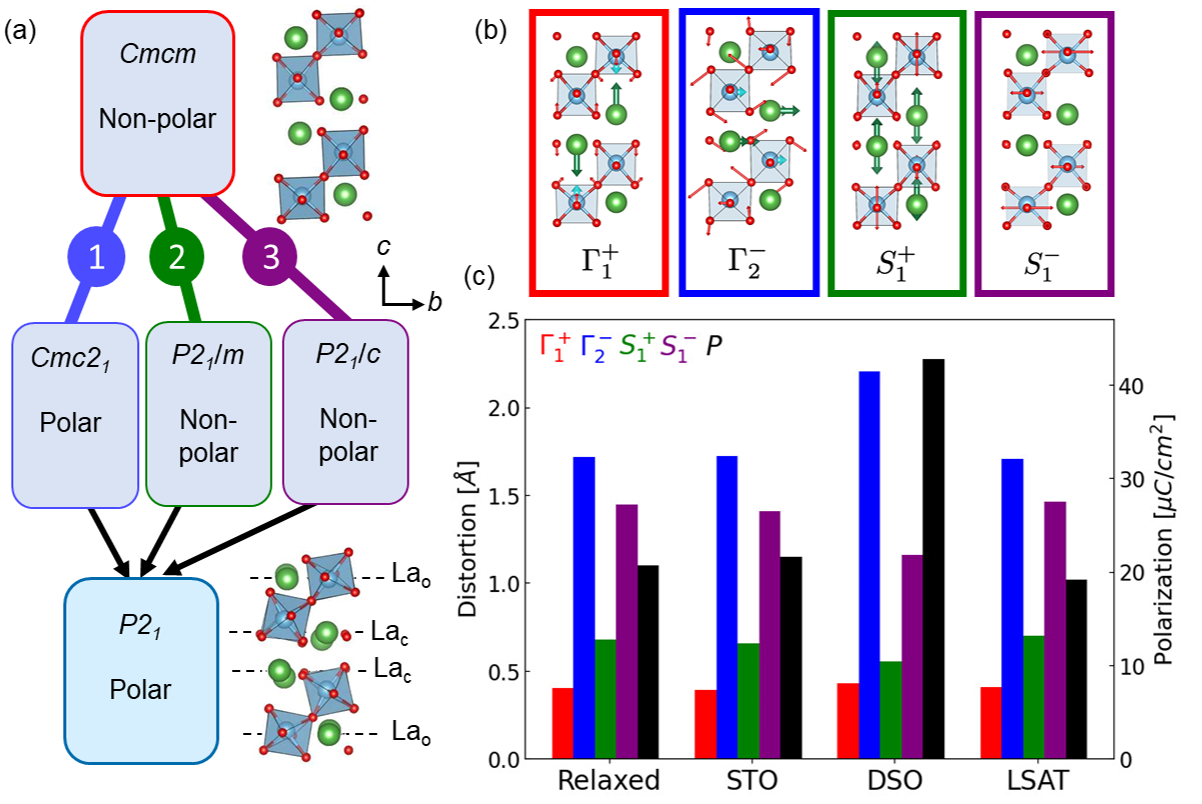}
    \caption{\textbf{Distortion modes in LTO and their dependence on epitaxial strain.} (a) Schematic of potential phase transitions from high-symmetry non-polar parent phase \(Cmcm\) to ferroelectric phase \(P2_1\) through pathways 1-3. In the polar \(P2_1\) unit cell, we can identify two La positions: the central La ions closest to the oxygen spacer (La\textsubscript{c}) and outer La ions (La\textsubscript{o}).   (b) Illustration of distortion modes connecting \(Cmcm\) and \(P2_1\) space groups. (c) Comparison of distortion amplitudes across different substrates indicating resilience to epitaxial strain with most pronounced changes in the amplitudes of \(S_1^+\), \(S_1^-\), and polar \(\Gamma_2^-\) modes. Black bars correspond to the resulting polarization values calculated with DFT, see Supporting Note S5.}
    \label{fig:dft}
\end{figure}

Given the strong interplay between strain and polar displacements in ferroelectrics, we turn to DFT calculations to gain deeper insight into the strain-dependent stability of the CG phase observed experimentally. Specifically, we explore how symmetry-adapted distortions within the LTO unit cell respond to varying epitaxial strain. The low-temperature ferroelectric phase of LTO is widely accepted as $P2_1$, with the high-symmetry phase at ca.\ 1500\textdegree{}C\cite{nanamatsuNewFerroelectricLa2Ti2o71974} as $Cmcm$\cite{lopez-perezInitioStudyProper2011, nunez_valdez_origin_2019, nanamatsuNewFerroelectricLa2Ti2o71974}, even though the transition pathway between them remains debated \cite{ishizawa_compounds_1982,nunez_valdez_origin_2019,ishizawaStructuralEvolutionTi2019}. The group-subgroup relationship between $Cmcm$ and $P2_1$ suggests two sequential phase transitions, and by performing a symmetry analysis we find four symmetry-adapted distortions connecting them. The first mode is the ferroelectric mode, $\Gamma_{2}^-$, which consists of a rotation of the octahedra along the $c$ axis and at the same time a lateral movement of central lanthanum (La\textsubscript{c}) ions, leading to a non-zero dipole moment along the \textit{b}-axis. This mechanism of inducing polarization was dubbed topological ferroelectricity\cite{lopez-perezInitioStudyProper2011}. Additionally, we identify two unit-cell doubling distortions, $S_1^+$ (octahedral tilts along $b$ and anti-polar La shifts) and $S_1^-$ (octahedral tilts along $c$), see Figure \ref{fig:dft}b. These three symmetry-breaking distortions couple to the non-symmetry-breaking $\Gamma_1^+$ mode, which contains anti-polar shifts of La\textsubscript{c} atoms and minor TiO\textsubscript{6} distortions.

This combination of distortions suggests three possible phase transition pathways. Two of these, $Cmcm \rightarrow Cmc2_1 \rightarrow P2_1$\cite{lopez-perezInitioStudyProper2011} and $Cmcm \rightarrow P2_1/m \rightarrow P2_1$\cite{nunez_valdez_origin_2019} have been previously discussed in the literature and are labeled as paths 1 and 2 in Figure \ref{fig:dft}a. We propose a third, previously unexplored sequence, $Cmcm \rightarrow P2_1/c \rightarrow P2_1$, labeled as path 3, in which the $S_1^-$ distortion mediates the intermediate phase. 
Our calculations reveal that the primary distortions are the $S_1^-$ tilts and the polar $\Gamma_2^-$ displacements (see Figure \ref{fig:dft}c), together contributing ca.\ 75\% of the total displacements. This suggests that these two modes likely act as primary order parameters in the phase transition, favoring either pathway 1 or 3. While some experimental evidence indicates $Cmc2_1$ as the intermediate phase\cite{ishizawa_compounds_1982}, the instability of the $S_1^-$ phonon, along with its flat phonon band (see Supporting Note S4), suggests that pathway 3 may be linked to experimentally observed incommensurate intermediate phases\cite{ishizawaStructuralEvolutionTi2019}. Ultimately, the exact sequence may depend on synthesis conditions, as was observed for LaTaO$_4$\cite{Howieson2021}, the \textit{n}=2 member of the CG series.

After identifying the main irreducible representations in the ferroelectric phase, we examined their response to the epitaxial strain imposed by each of the substrates. Across the three substrates studied, we found that the distortions are generally resilient to considerable strain values (see Figure \ref{fig:dft}c). STO, which provides the closest epitaxial match, shows distortion amplitudes nearly identical to those in the relaxed structure. Similarly, LSAT, which introduces about 1\% compressive strain along $a$ and $b$, produces only minor changes in distortion amplitude. This resilience is unusual but aligns with LTO’s high ferroelectric T\textsubscript{c}, underscoring the material’s robust ferroelectric state. The effect is notably different on DSO, which imposes a substantial tensile strain of almost 3\% along the $b$ direction and 1\% along $a$. While the overall distortion amplitude on DSO is only moderately increased relative to relaxed LTO, a significant redistribution among the distortion modes occurs: the polar $\Gamma_2^-$ mode rises by approximately 30\%, while $S_1^+$ and $S_1^-$ tilt patterns decrease slightly. These observations suggest that tensile strain in LTO significantly enhances its in-plane polarization.

Motivated by this finding, we calculate the spontaneous polarization for the different strain states (see Figure \ref{fig:dft}c and Supporting Note S5), and find spontaneous polarization of 20.7 \textmu C/cm\textsuperscript{2} for the relaxed structure, comparable to the literature value\cite{nunez_valdez_origin_2019}. Under epitaxial strain, we obtain 21.6 \textmu C/cm\textsuperscript{2} and 19.2 \textmu C/cm\textsuperscript{2} for STO and LSAT respectively. The most striking result is observed with DSO, where the polarization value doubles to 42.7 \textmu C/cm\textsuperscript{2}, suggesting a strong non-linear behavior of the mode effective charges under strain for the $\Gamma_2^-$ distortion. This marked increase in LTO polarization under tensile strain likely drives the stabilization of the CG phase over the perovskite or $\gamma$-polymorph phases, explaining its distinctive stability on the DSO substrate observed experimentally.

\subsection{Achieving CG phase stability across substrates via post-annealing: structural characterization and mapping of atomic displacements}

\begin{figure}[t!]
  \centering 
  \begin{adjustbox}{width=0.88\textwidth, center}
    \includegraphics[width=0.88\textwidth, clip, trim=4 15 4 4]{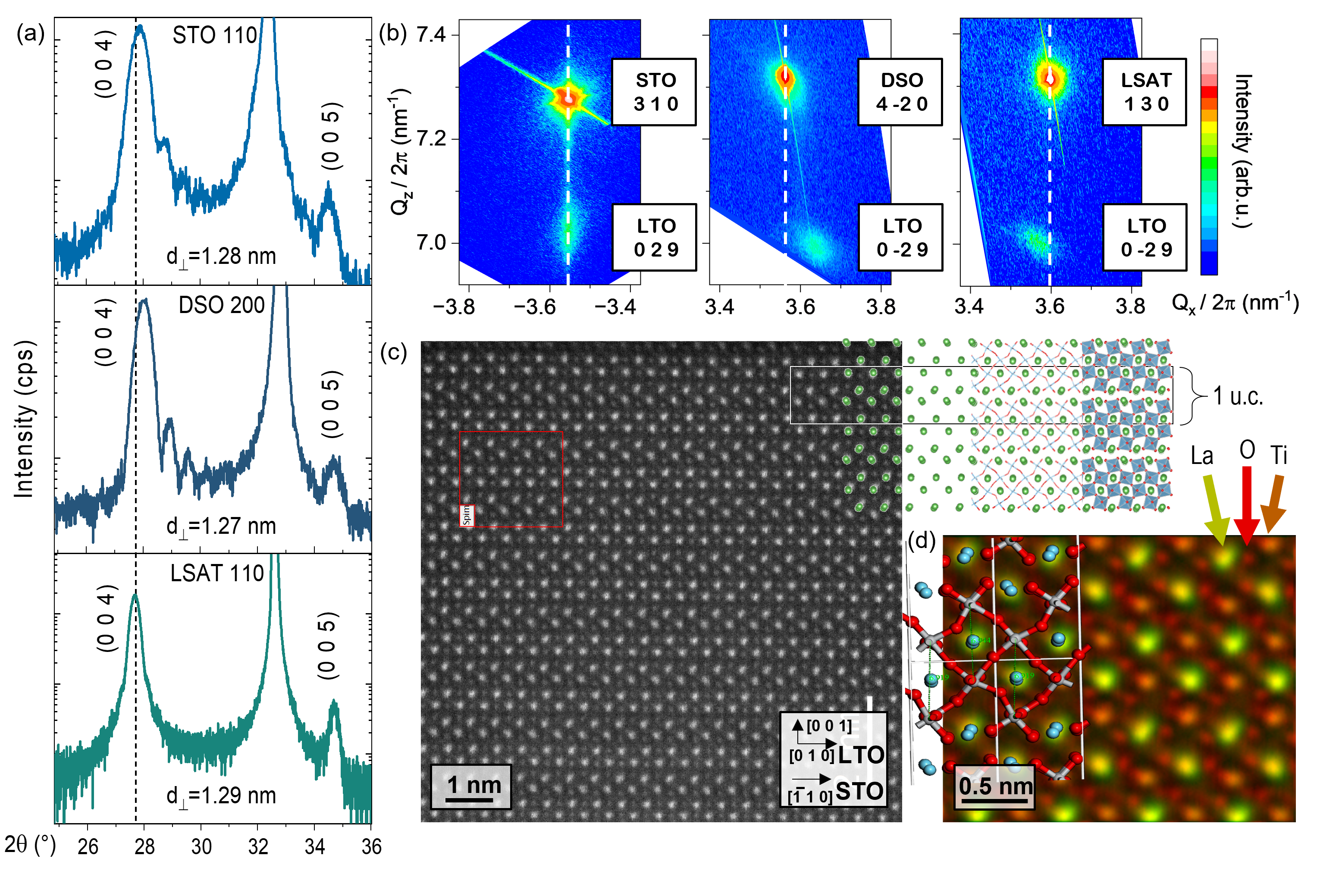}
  \end{adjustbox}
  \caption{\textbf{Structural properties of LTO films of the CG phase: post-annealed on STO and LSAT and as-grown on DSO.} (a) Diffractograms of LTO films show the stabilized 001-oriented CG phase. Strain effects are evident from the shifts of the 004 reflections around the literature value (dashed line). (b) Reciprocal space maps of LTO (0 2 9) and (0 -2 9) reflections reveal coherently strained films on the STO substrate, with relaxation observed on DSO and LSAT. (c) HAADF-STEM image of the LTO film on STO (110) displays the precise atomic layering in the \textit{b}-\textit{c} plane, highlighting the alternation of La ions' zig-zag rows, indicative of polar distortions. The inset in the top right provides the overlay with the theoretical atomic positions, bonds, and oxygen octahedra. (d) 4D-STEM analysis (area marked with a red box in (c)) with superimposed theoretical atomic positions of La, Ti, and O atoms.}\label{fig:struct_properties}
\end{figure}

While the tensile strain of the DSO substrate is required to stabilize the CG phase during the deposition through the controlled layer-by-layer growth mode, literature reports indicate that annealing LTO films in an oxygen atmosphere at temperatures above 1100\textdegree{}C can serve as a post-deposition method for obtaining the CG phase\cite{nezu_solid-phase_2017,kasparTuningPiezoelectricProperties2018, Yao2020, qiaoMicrostructureElectricalProperties2022}. This strategy indeed proved to be effective for rectifying the correct CG layering on the STO and LSAT substrates, which fail to stabilize it in the pristine films. We stress here, however, that such thermal intervention comes at the expense of losing detailed control over structural properties such as surface morphology and exact RHEED-monitored thickness. Figure \ref{fig:struct_properties}a displays the diffractograms of the post-annealed LTO films on STO and LSAT substrates alongside the pristine LTO film on DSO: all three films now exhibit the 001-oriented CG phase, free from detectable impurities (see Supporting Note S6) and crystal twinning. Reciprocal space mapping of LTO (0 2 9) and (0 -2 9) reflections shown in Figure \ref{fig:struct_properties}b reveals that the films are coherently strained to the STO substrate, while strain relaxation is observed on both DSO and LSAT. Despite not being fully strained to their substrates, these films retain characteristics of the tensile and compressive strains applied. For example, shifts in the 004 reflections relative to the bulk out-of-plane lattice parameter\cite{schmalle_twin_1993} (marked with a dashed line, corresponding to d\textsubscript{$\perp$}=1.2864 nm) are evident in Figure \ref{fig:struct_properties}a, where the LTO out-of-plane lattice parameter is elongated by 0.2\% on LSAT and shortened by -0.3\% and -1.0\% on STO and DSO, respectively, consistent with the expected strain effects.

To complement the macroscopic structural analysis, we conducted atomically-resolved high-angle annular dark-field scanning transmission electron microscopy (HAADF-STEM) imaging of the LTO films on STO (110). The image displayed in Figure \ref{fig:struct_properties}c shows the characteristic layering of atomic planes of the CG phase in the \textit{b}-\textit{c} plane of the film. La ions have the brightest contrast and form characteristic zig-zag-like slabs: with and without teardrop shape distortions in La, that alternate every two atomic rows along the out-of-plane [001]\textsubscript{LTO} direction. The distortions, seen in La ions closest to the oxygen spacers (marked La\textsubscript{c} in Figure \ref{fig:dft}a), arise from the $\Gamma^{-}_{2}$ polar distortion mode (shown in Figure \ref{fig:dft}b), which induces lateral movement of La ions along the \textit{b}-axis. In combination with the anti-polar ($\Gamma_1^+$ and $S1^+$) movement along the $c$ axis, this leads to the appearance of the teardrop-like La doublets. In contrast, the two outer La ions, positioned further from the oxygen planes (marked La\textsubscript{o} in Figure \ref{fig:dft}a), exhibit no in-plane displacement and therefore appear perfectly spherical and one can only observe the lateral movement of Ti atoms positioned above and below them, again consistent with the $\Gamma^{-}_{2}$ mode. The inset in the top right provides the overlay with the theoretical atomic positions, bonds, and oxygen octahedra associated with the monoclinic space group \textit{$P2_1$}\cite{schmalle_twin_1993}. A selected portion of the film (red box) is analyzed with 4D-STEM to identify the exact positions of La, oxygen (O), and titanium (Ti) ions (Figure \ref{fig:struct_properties}d).

\subsection{Significantly enhanced ferroelectricity with no critical thickness, ferroelectric domains and their formation mechanisms in post-annealed LTO films}

\begin{figure}[t!]
  \centering 
  \begin{adjustbox}{width=0.88\textwidth, center}
    \includegraphics[width=0.88\textwidth,clip, trim=10 10 3 5]{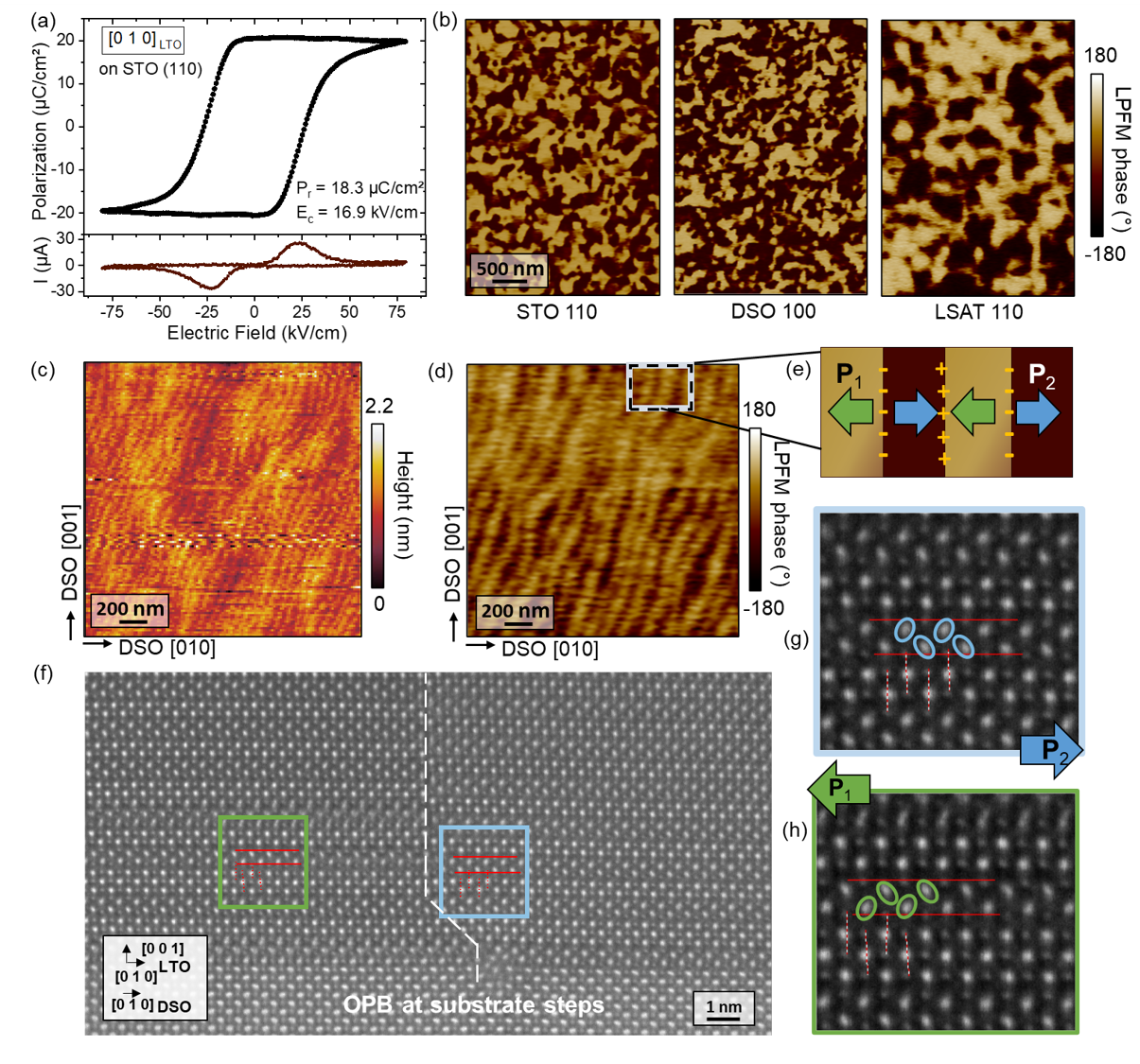}
  \end{adjustbox}
  \caption{\textbf{Ferroelectric properties of the LTO films.} (a) Polarization-Electric Field (P-E) hysteresis loops and switching current recorded for a 56-nm-thick LTO film on STO (110), displaying clear ferroelectric behavior along the polar axis \hbox{[0 1 0]\textsubscript{LTO}}. (b) LPFM phase images showing in-plane ferroelectric domains in ca. 20-unit-cell-thick LTO films on STO, DSO, and LSAT. The images reveal two domain variants with polarizations along [0 1 0]\textsubscript{LTO} and \mbox{[0 $\bar1$ 0]\textsubscript{LTO}}. (c) Topography of a monolayer LTO film on DSO with atomically flat terraces. (d) The corresponding PFM image shows in-plane-polarized stripe domains, confirming the ferroelectric functionality of just a 1.3-nm-thick LTO layer. (e) Schematic of the stripe domain pattern giving rise to nominally charged domain walls with alternating accumulations of mobile screening charges. (f) HAADF-STEM imaging of the LTO film adjacent to a DSO substrate step reveals an out-of-phase boundary (OPB) in the film. Atomically resolved displacements (g) to the right and (h) to the left of the OPB show opposite in-plane polarization directions. This suggests the presence of a tail-to-tail domain wall pinned at the substrate step position.}\label{fig:ferro}
\end{figure}

After confirming the structural quality of our LTO films on all three substrates, we proceeded to further characterization of their ferroelectric properties. Figure \ref{fig:ferro}a illustrates the ferroelectric switching characteristics of a representative 56-nm-thick LTO film on STO (110) measured using a Positive Up Negative Down (PUND) pulse scheme. The field is applied in the plane of the film along the polar axis [0 1 0]\textsubscript{LTO}. We obtain a clear ferroelectric polarization hysteresis loop in line with the expected uniaxial polar anisotropy observed in vector PFM measurements (refer to Supporting Note S3). The measured polarization of 18.3 \textmu C/cm\textsuperscript{2} is nearly four times higher than 5 \textmu C/cm\textsuperscript{2} reported for bulk LTO crystals\cite{nanamatsuNewFerroelectricLa2Ti2o71974} or 2.8 \textmu C/cm\textsuperscript{2} measured for LTO films on Nb-doped STO (110)\cite{shao_microstructure_2011}. Yet, our polarization value is not too far off from the polarization of approx. 21 \textmu C/cm\textsuperscript{2} predicted from density functional theory calculations (see section 2.3). This substantial enhancement in polarization emphasizes the importance of epitaxially oriented films with improved crystal quality and epitaxially enhanced polar displacements. Furthermore, the coercive field of 16.9 kV/cm is more than three times lower than the previously reported value\cite{nanamatsu_crystallographic_1975}, rendering such LTO films even more attractive for device applications as their operational voltage could be significantly reduced.

Building on the insights gained from the microscopic polarization analysis, we now turn our attention to the PFM-based ferroelectric domain analysis. While there have been reports on the PFM measurements performed on the LTO films in the literature, they were obscured by significant surface roughness creating a non-negligible cross-talk between topography and piezoresponse-related tip torsion and buckling. In our case, the achievement of atomically flat films with uniaxial in-plane polarization along their \textit{b}-axis facilitates the clear identification of ferroelectric domains in CG films. Figure \ref{fig:ferro}b displays LPFM phase images of ferroelectric in-plane domains recorded for ca. 20-unit-cell-thick LTO films on the three different substrates. Across all three strain states, the patterns contain the two domain variants with polarization pointing along [0 1 0]\textsubscript{LTO} and \mbox{[0 $\bar1$ 0]\textsubscript{LTO}}. We noted two general trends for domain scaling: ferroelectric domain size increases with both film thickness and compressive strain.

In-plane-polarized ferroelectric films should, in principle, exhibit no critical thickness for ferroelectricity as the resulting bound charges are not localized on the film surfaces but rather on domain walls within the film. The stabilized layer-by-layer growth mode of LTO films on DSO gives us the means to prepare ultrathin films of just 1 u.c. in thickness and test this. Figure \ref{fig:ferro}c displays AFM topography of a 1-unit-cell-thick LTO coverage, maintaining atomic step terraces of the underlying DSO substrate. The corresponding LPFM image (Figure \ref{fig:ferro}d) reveals well-ordered stripe domains extending over hundreds of microns, confirming that a mere 1.3-nm-thick layer of LTO already exhibits its ferroelectric functionality. As the polarization is along the \textit{b}-axis, such a domain pattern can be described as a network of partially charged domain walls with alternating accumulations of screening mobile charges, as schematized in Figure \ref{fig:struct_properties}e. The similarity between the domain shape and the atomic terraces in the topography (cf.\ Figure \ref{fig:ferro}c) suggests that structural defects at each substrate step likely trigger the reversal of in-plane polarization from one domain to the next. This hypothesis is supported by HAADF-STEM imaging of the LTO film near a DSO substrate step (Figure \ref{fig:ferro}f), where the substrate step induces an out-of-register shift between the LTO unit cells, known as an out-of-phase boundary (OPB)\cite{Zurbuchen2007}. Analysis of the atomic displacements in the film to the right of the OPB (marked by a blue box and zoomed in in Figure \ref{fig:ferro}g) shows that the teardrop-like La displacements point left, while the non-distorted La columns are displaced right relative to the Ti-Ti line, indicating rightward polarization. Conversely, to the left of the OPB (marked by a green box and zoomed in Figure \ref{fig:ferro}h), the direction of the teardrop displacements reverses, pointing right, while the non-distorted La columns shift left, confirming the inversion of polarization direction and the formation of a tail-to-tail domain wall at the OPB. This domain-formation mechanism, similar to that observed in layered in-plane-polarized Aurivillius systems\cite{Gradauskaite2022a, keeney_what_2023}, arises from discontinuities in electrostatic boundary conditions due to structural defects. It is likely universally applicable across all layered compounds, such as Ruddlesden-Popper or Dion-Jacobson phases, when they are epitaxially stabilized as uniaxial in-plane ferroelectrics.

\section{Conclusion} 
In this work, we systematically investigated the epitaxial growth of LTO films and the influence of various strain states on the stability of the layered CG phase. We demonstrate that high-quality films exhibit robust ferroelectric polarization around 18 \textmu C/cm\textsuperscript{2} — more than three times higher than previously reported, aligning with theoretical predictions. Our study reveals a remarkable resilience of the ferroelectric phase under epitaxial strain. Specifically, a tensile strain of 3\% along the polar axis, introduced by DSO (100) substrates, not only preserves but enhances the CG phase, promoting a coalescent layer-by-layer growth mode. In contrast, STO (110) and LSAT (110) substrates, exerting negligible or compressive strain, initially fail to stabilize the CG phase but can achieve phase stability through post-deposition annealing. Complementing experimental findings, our DFT calculations show that the CG phase maintains structural stability under varying strains, even predicting a doubling of polarization on DSO substrates. Lastly, we demonstrate that ferroelectricity is retained down to a single unit cell, with stripe-domain patterns indicating a complex interplay between substrate imperfections and domain configurations. These findings establish Carpy-Galy thin films as a promising platform for exploring ultrathin ferroelectric applications. Our work invites further research into novel functionalities in this layered family, which could expand the capabilities of next-generation electronic devices.
\newpage

\section{Experimental Section}
\RaggedRight

\subsection*{Thin-Film Growth}\justifying La\textsubscript{2}Ti\textsubscript{2}O\textsubscript{7} (LTO) films were grown on SrTiO\textsubscript{3} (STO) (110) (Surfacenet GmbH), DyScO\textsubscript{3} (DSO) (100) (Surfacenet GmbH), and LaAlO\textsubscript{3}-Sr\textsubscript{2}TaAlO\textsubscript{6} (LSAT) (110) substrates (CrysTec GmbH). The STO substrates underwent chemical treatment with a buffered HF solution, and all substrates were thermally treated to form well-defined atomic step terraces\cite{biswas_atomically_2017}. The films were deposited using a single stoichiometric La\textsubscript{2}Ti\textsubscript{2}O\textsubscript{7} target by pulsed laser deposition using a 248 nm KrF excimer laser, at substrate temperatures of 800\textdegree{}C to 1100\textdegree{}C, oxygen partial pressure of 5$\cdot$10\textsuperscript{$-$3} to 6$\cdot$10\textsuperscript{$-$2} mbar, laser fluence of 1.0 to 1.4 J cm\textsuperscript{$-$2} with a laser spot size of 3.5 $\pm$ 0.3 mm\textsuperscript{2} and \mbox{1 Hz} repetition rate. 

Optimized parameters for each substrate were as follows: on STO (110): 800\textdegree{}C, 6$\cdot$10\textsuperscript{$-$2} mbar, 1.2 J cm\textsuperscript{$-$2}, followed by post-annealing in an oxygen atmosphere at 1100\textdegree{}C for 3 hours; on DSO (100): 1075\textdegree{}C, 7.5$\cdot$10\textsuperscript{$-$3} mbar, 1.4 J cm\textsuperscript{$-$2}, with no post-annealing required; on LSAT (110): 1075\textdegree{}C, 5$\cdot$10\textsuperscript{$-$3} mbar, 1.2 J cm\textsuperscript{$-$2}, followed by post-annealing in an oxygen atmosphere at 1100\textdegree{}C for 3 hours. The thickness of the films was continuously monitored using RHEED during growth and verified post-deposition by X-ray reflectivity.

\subsection*{X-ray Diffraction}\justifying  The crystalline structure of thin films was analyzed by X-ray diffraction, X-ray reflectivity, and reciprocal space mapping measurements using a four-circle X-ray diffractometer (Panalytical Empyrean and Rigaku SmartLab).

\subsection*{Scanning Probe Microscopy}\justifying  Atomic and piezoresponse force microscopy acquisition was performed in contact mode in a Nanoscope V multimode (Bruker) microscope with two external SR830 lock-in amplifiers (Stanford Research) for simultaneous acquisition of in-plane and out-of-plane piezoresponse. The data acquisition was performed using AC modulation of 5-12 V (peak-to-peak) at 35 kHz applied to the Pt-coated tip (2.7 N/m cantilever, ElectriAll-In-One). Images of in-plane-polarized domains were recorded through cantilever torsion mode when measuring with a cantilever perpendicular to the uniaxial LTO polarization axis [0 1 0]\textsubscript{LTO}. 

 \subsection*{Interdigitated Electrode Design}\justifying  The interdigitated electrodes were patterned using maskless projection photolithography on a Smart Print UV system (microlight3D). This was followed by the sputtering of a 3 nm titanium (Ti) adhesion layer and a 10 to 20 nm gold (Au) layer. The patterning process was completed with a lift-off process. The electrode pattern comprises 200 fingers, each with a width and gap of 5 µm, and a length of 995 \SI{}{\micro\metre} each. Sets of interdigitated electrodes were designed on the films to probe in-plane polarization along the [0 1 0]\textsubscript{LTO} crystallographic axis. 
 
 \subsection*{Macroscopic Ferroelectric Polarization Switching}\justifying  Ferroelectric properties of the LTO films were characterized with a ferroelectric tester (Radiant Technologies). The \textit{P–E} loops were measured using the positive-up negative-down (PUND) method. This technique involved applying a sequence of five electrical pulses over a total period of 1 ms, which allowed for precise evaluation of the intrinsic switching properties of the ferroelectric films, free from extrinsic contributions such as leakage current.

 \subsection*{Scanning Transmission Electron Microscopy (STEM)}\justifying  STEM experiments were performed using a Cs-corrected Nion USTEM operated at 200 keV, with a probe current of approximately 20 pA and a probe size of about 80 pm. HAADF images were obtained with typical 80-200 mrad collection angles. 4D-STEM images were also obtained by collecting the entire diffraction patterns, with a typical spatial step of 10 to 20 pm. Images were then generated (e.g., Figure \ref{fig:struct_properties}d) by computing the divergence of the center-of-mass of the diffraction patterns, which roughly approximate charge distribution images. Samples were prepared for STEM using a focused ion beam, with a Ga beam at 30, 16, and 8 keV, followed by final surface cleaning at 5 and 2 keV.

\subsection*{Density Functional Theory (DFT) Calculations}\justifying  
  DFT calculations were performed using the \textsc{VASP}\cite{kresseInitioMolecularDynamics1993,kresseEfficiencyInitioTotal1996} code. A converged \(6 \times 6 \times 2\) grid is used for the 44 atom \(P2_1\) unit cell and \(6 \times 6 \times 6\) for the 22 atom \(Cmcm\) cell. A planewave cutoff of 500 eV is employed. The pseudopotentials are used with the following electrons in the valence: for La: \(5s^2 5p^6 6s^2 5d^1\); for Ti: \(4s^2 3d^2\); and for O: \(2s^2 2p^4\). In order to calculate the adjustment of the material to the substrate, the \(a\) and \(b\) lattice parameters are fixed, and the relaxation of the out-of-plane lattice parameter and the monoclinic angle between the \(a\) and \(c\) axes is allowed. Internal forces were relaxed down to a maximal force of 0.01 eV/$\AA$. Symmetry mode analysis comparing $Cmcm$ and $P2_1$ was performed using the \textsc{AMPLIMODES} sofware\cite{orobengoaAMPLIMODESSymmetrymodeAnalysis2009}.

\section*{Acknowledgements}
E.G. acknowledges the Swiss National Science Foundation for financial support under Project No.\ P500PT\_214449. M.B. and A.S.G. acknowledge ERC for funding under PoC UPLIFT Project No.\ 101113273. The authors thank Luis Moreno Vicente-Arche for experimental support.

\section*{Conflict of Interest} The authors declare no conflict of interest.

\section*{Author contributions}
The thin-film growth, scanning probe microscopy, and XRD structural analysis were conducted by E.G..  Interdigitated top electrodes were designed and patterned by A.S.G. with E.G.. Ferroelectric switching experiments were performed by E.G. with A.S.G.. X.L. and A.G. carried out the STEM investigations. L.I. supported the thin-film growth and XRD experiments. DFT calculations were performed by Q.N.M.. The experiment was designed by E.G. with M.B., who also supervised the study. All authors contributed to the discussion of the results. E.G. wrote the manuscript with the inputs from all authors. 


\bibliographystyle{apsrev4-2}
\bibliography{Carpy-Galy, Aurivillius}

\begin{thebibliography}{49}%
\makeatletter
\providecommand \@ifxundefined [1]{%
 \@ifx{#1\undefined}
}%
\providecommand \@ifnum [1]{%
 \ifnum #1\expandafter \@firstoftwo
 \else \expandafter \@secondoftwo
 \fi
}%
\providecommand \@ifx [1]{%
 \ifx #1\expandafter \@firstoftwo
 \else \expandafter \@secondoftwo
 \fi
}%
\providecommand \natexlab [1]{#1}%
\providecommand \enquote  [1]{``#1''}%
\providecommand \bibnamefont  [1]{#1}%
\providecommand \bibfnamefont [1]{#1}%
\providecommand \citenamefont [1]{#1}%
\providecommand \href@noop [0]{\@secondoftwo}%
\providecommand \href [0]{\begingroup \@sanitize@url \@href}%
\providecommand \@href[1]{\@@startlink{#1}\@@href}%
\providecommand \@@href[1]{\endgroup#1\@@endlink}%
\providecommand \@sanitize@url [0]{\catcode `\\12\catcode `\$12\catcode `\&12\catcode `\#12\catcode `\^12\catcode `\_12\catcode `\%12\relax}%
\providecommand \@@startlink[1]{}%
\providecommand \@@endlink[0]{}%
\providecommand \url  [0]{\begingroup\@sanitize@url \@url }%
\providecommand \@url [1]{\endgroup\@href {#1}{\urlprefix }}%
\providecommand \urlprefix  [0]{URL }%
\providecommand \Eprint [0]{\href }%
\providecommand \doibase [0]{https://doi.org/}%
\providecommand \selectlanguage [0]{\@gobble}%
\providecommand \bibinfo  [0]{\@secondoftwo}%
\providecommand \bibfield  [0]{\@secondoftwo}%
\providecommand \translation [1]{[#1]}%
\providecommand \BibitemOpen [0]{}%
\providecommand \bibitemStop [0]{}%
\providecommand \bibitemNoStop [0]{.\EOS\space}%
\providecommand \EOS [0]{\spacefactor3000\relax}%
\providecommand \BibitemShut  [1]{\csname bibitem#1\endcsname}%
\let\auto@bib@innerbib\@empty
\bibitem [{\citenamefont {Junquera}\ and\ \citenamefont {Ghosez}(2003)}]{Junquera2003}%
  \BibitemOpen
  \bibfield  {author} {\bibinfo {author} {\bibfnamefont {J.}~\bibnamefont {Junquera}}\ and\ \bibinfo {author} {\bibfnamefont {P.}~\bibnamefont {Ghosez}},\ }\href {https://doi.org/10.1038/nature01501} {\bibfield  {journal} {\bibinfo  {journal} {Nature}\ }\textbf {\bibinfo {volume} {422}},\ \bibinfo {pages} {506} (\bibinfo {year} {2003})}\BibitemShut {NoStop}%
\bibitem [{\citenamefont {Li}\ \emph {et~al.}(2020)\citenamefont {Li}, \citenamefont {Tan},\ and\ \citenamefont {Duan}}]{li_hexagonal_2020}%
  \BibitemOpen
  \bibfield  {author} {\bibinfo {author} {\bibfnamefont {M.}~\bibnamefont {Li}}, \bibinfo {author} {\bibfnamefont {H.}~\bibnamefont {Tan}},\ and\ \bibinfo {author} {\bibfnamefont {W.}~\bibnamefont {Duan}},\ }\href {https://doi.org/10.1039/d0cp02195d} {\bibfield  {journal} {\bibinfo  {journal} {Physical Chemistry Chemical Physics}\ }\textbf {\bibinfo {volume} {22}},\ \bibinfo {pages} {14415} (\bibinfo {year} {2020})}\BibitemShut {NoStop}%
\bibitem [{\citenamefont {Bousquet}\ \emph {et~al.}(2008)\citenamefont {Bousquet}, \citenamefont {Dawber}, \citenamefont {Stucki}, \citenamefont {Lichtensteiger}, \citenamefont {Hermet}, \citenamefont {Gariglio}, \citenamefont {Triscone},\ and\ \citenamefont {Ghosez}}]{Bousquet2008}%
  \BibitemOpen
  \bibfield  {author} {\bibinfo {author} {\bibfnamefont {E.}~\bibnamefont {Bousquet}}, \bibinfo {author} {\bibfnamefont {M.}~\bibnamefont {Dawber}}, \bibinfo {author} {\bibfnamefont {N.}~\bibnamefont {Stucki}}, \bibinfo {author} {\bibfnamefont {C.}~\bibnamefont {Lichtensteiger}}, \bibinfo {author} {\bibfnamefont {P.}~\bibnamefont {Hermet}}, \bibinfo {author} {\bibfnamefont {S.}~\bibnamefont {Gariglio}}, \bibinfo {author} {\bibfnamefont {J.~M.}\ \bibnamefont {Triscone}},\ and\ \bibinfo {author} {\bibfnamefont {P.}~\bibnamefont {Ghosez}},\ }\href {https://doi.org/10.1038/nature06817} {\bibfield  {journal} {\bibinfo  {journal} {Nature}\ }\textbf {\bibinfo {volume} {452}},\ \bibinfo {pages} {732} (\bibinfo {year} {2008})}\BibitemShut {NoStop}%
\bibitem [{\citenamefont {Park}\ \emph {et~al.}(2018)\citenamefont {Park}, \citenamefont {Lee}, \citenamefont {Mikolajick}, \citenamefont {Schroeder},\ and\ \citenamefont {Hwang}}]{Park2018}%
  \BibitemOpen
  \bibfield  {author} {\bibinfo {author} {\bibfnamefont {M.~H.}\ \bibnamefont {Park}}, \bibinfo {author} {\bibfnamefont {Y.~H.}\ \bibnamefont {Lee}}, \bibinfo {author} {\bibfnamefont {T.}~\bibnamefont {Mikolajick}}, \bibinfo {author} {\bibfnamefont {U.}~\bibnamefont {Schroeder}},\ and\ \bibinfo {author} {\bibfnamefont {C.~S.}\ \bibnamefont {Hwang}},\ }\href {https://doi.org/10.1557/mrc.2018.175} {\bibfield  {journal} {\bibinfo  {journal} {MRS Communications}\ }\textbf {\bibinfo {volume} {8}},\ \bibinfo {pages} {795} (\bibinfo {year} {2018})}\BibitemShut {NoStop}%
\bibitem [{\citenamefont {Guan}\ \emph {et~al.}(2020)\citenamefont {Guan}, \citenamefont {Hu}, \citenamefont {Shen}, \citenamefont {Xiang}, \citenamefont {Zhong}, \citenamefont {Chu},\ and\ \citenamefont {Duan}}]{Guan2020}%
  \BibitemOpen
  \bibfield  {author} {\bibinfo {author} {\bibfnamefont {Z.}~\bibnamefont {Guan}}, \bibinfo {author} {\bibfnamefont {H.}~\bibnamefont {Hu}}, \bibinfo {author} {\bibfnamefont {X.}~\bibnamefont {Shen}}, \bibinfo {author} {\bibfnamefont {P.}~\bibnamefont {Xiang}}, \bibinfo {author} {\bibfnamefont {N.}~\bibnamefont {Zhong}}, \bibinfo {author} {\bibfnamefont {J.}~\bibnamefont {Chu}},\ and\ \bibinfo {author} {\bibfnamefont {C.}~\bibnamefont {Duan}},\ }\href {https://doi.org/10.1002/aelm.201900818} {\bibfield  {journal} {\bibinfo  {journal} {Advanced Electronic Materials}\ }\textbf {\bibinfo {volume} {6}},\ \bibinfo {pages} {1900818} (\bibinfo {year} {2020})}\BibitemShut {NoStop}%
\bibitem [{\citenamefont {Xu}\ \emph {et~al.}(2019)\citenamefont {Xu}, \citenamefont {Kopyl}, \citenamefont {Kholkin},\ and\ \citenamefont {Rocha}}]{xu_hybrid_2019}%
  \BibitemOpen
  \bibfield  {author} {\bibinfo {author} {\bibfnamefont {W.~J.}\ \bibnamefont {Xu}}, \bibinfo {author} {\bibfnamefont {S.}~\bibnamefont {Kopyl}}, \bibinfo {author} {\bibfnamefont {A.}~\bibnamefont {Kholkin}},\ and\ \bibinfo {author} {\bibfnamefont {J.}~\bibnamefont {Rocha}},\ }\href {https://doi.org/10.1016/j.ccr.2019.02.012} {\bibfield  {journal} {\bibinfo  {journal} {Coordination Chemistry Reviews}\ }\textbf {\bibinfo {volume} {387}},\ \bibinfo {pages} {398} (\bibinfo {year} {2019})}\BibitemShut {NoStop}%
\bibitem [{\citenamefont {Benedek}\ \emph {et~al.}(2015)\citenamefont {Benedek}, \citenamefont {Rondinelli}, \citenamefont {Djani}, \citenamefont {Ghosez},\ and\ \citenamefont {Lightfoot}}]{benedek_understanding_2015}%
  \BibitemOpen
  \bibfield  {author} {\bibinfo {author} {\bibfnamefont {N.~A.}\ \bibnamefont {Benedek}}, \bibinfo {author} {\bibfnamefont {J.~M.}\ \bibnamefont {Rondinelli}}, \bibinfo {author} {\bibfnamefont {H.}~\bibnamefont {Djani}}, \bibinfo {author} {\bibfnamefont {P.}~\bibnamefont {Ghosez}},\ and\ \bibinfo {author} {\bibfnamefont {P.}~\bibnamefont {Lightfoot}},\ }\href {https://doi.org/10.1039/C5DT00010F} {\bibfield  {journal} {\bibinfo  {journal} {Dalton Transactions}\ }\textbf {\bibinfo {volume} {44}},\ \bibinfo {pages} {10543} (\bibinfo {year} {2015})}\BibitemShut {NoStop}%
\bibitem [{\citenamefont {Keeney}\ \emph {et~al.}(2020{\natexlab{a}})\citenamefont {Keeney}, \citenamefont {Saghi}, \citenamefont {O'Sullivan}, \citenamefont {Alaria}, \citenamefont {Schmidt},\ and\ \citenamefont {Colfer}}]{Keeney2020a}%
  \BibitemOpen
  \bibfield  {author} {\bibinfo {author} {\bibfnamefont {L.}~\bibnamefont {Keeney}}, \bibinfo {author} {\bibfnamefont {Z.}~\bibnamefont {Saghi}}, \bibinfo {author} {\bibfnamefont {M.}~\bibnamefont {O'Sullivan}}, \bibinfo {author} {\bibfnamefont {J.}~\bibnamefont {Alaria}}, \bibinfo {author} {\bibfnamefont {M.}~\bibnamefont {Schmidt}},\ and\ \bibinfo {author} {\bibfnamefont {L.}~\bibnamefont {Colfer}},\ }\href {https://doi.org/10.1021/acs.chemmater.0c03454} {\bibfield  {journal} {\bibinfo  {journal} {Chemistry of Materials}\ }\textbf {\bibinfo {volume} {32}},\ \bibinfo {pages} {10511} (\bibinfo {year} {2020}{\natexlab{a}})}\BibitemShut {NoStop}%
\bibitem [{\citenamefont {Gradauskaite}\ \emph {et~al.}(2020)\citenamefont {Gradauskaite}, \citenamefont {Campanini}, \citenamefont {Biswas}, \citenamefont {Schneider}, \citenamefont {Fiebig}, \citenamefont {Rossell},\ and\ \citenamefont {Trassin}}]{Gradauskaite2020a}%
  \BibitemOpen
  \bibfield  {author} {\bibinfo {author} {\bibfnamefont {E.}~\bibnamefont {Gradauskaite}}, \bibinfo {author} {\bibfnamefont {M.}~\bibnamefont {Campanini}}, \bibinfo {author} {\bibfnamefont {B.}~\bibnamefont {Biswas}}, \bibinfo {author} {\bibfnamefont {C.~W.}\ \bibnamefont {Schneider}}, \bibinfo {author} {\bibfnamefont {M.}~\bibnamefont {Fiebig}}, \bibinfo {author} {\bibfnamefont {M.~D.}\ \bibnamefont {Rossell}},\ and\ \bibinfo {author} {\bibfnamefont {M.}~\bibnamefont {Trassin}},\ }\href {https://doi.org/10.1002/admi.202000202} {\bibfield  {journal} {\bibinfo  {journal} {Advanced Materials Interfaces}\ }\textbf {\bibinfo {volume} {7}},\ \bibinfo {pages} {2000202} (\bibinfo {year} {2020})}\BibitemShut {NoStop}%
\bibitem [{\citenamefont {Gradauskaite}\ \emph {et~al.}(2021)\citenamefont {Gradauskaite}, \citenamefont {Gray}, \citenamefont {Campanini}, \citenamefont {Rossell},\ and\ \citenamefont {Trassin}}]{gradauskaite2021a}%
  \BibitemOpen
  \bibfield  {author} {\bibinfo {author} {\bibfnamefont {E.}~\bibnamefont {Gradauskaite}}, \bibinfo {author} {\bibfnamefont {N.}~\bibnamefont {Gray}}, \bibinfo {author} {\bibfnamefont {M.}~\bibnamefont {Campanini}}, \bibinfo {author} {\bibfnamefont {M.~D.}\ \bibnamefont {Rossell}},\ and\ \bibinfo {author} {\bibfnamefont {M.}~\bibnamefont {Trassin}},\ }\href {https://pubs.acs.org/doi/abs/10.1021/acs.chemmater.1c03466} {\bibfield  {journal} {\bibinfo  {journal} {Chemistry of Materials}\ }\textbf {\bibinfo {volume} {33}},\ \bibinfo {pages} {9439} (\bibinfo {year} {2021})}\BibitemShut {NoStop}%
\bibitem [{\citenamefont {Keeney}\ \emph {et~al.}(2020{\natexlab{b}})\citenamefont {Keeney}, \citenamefont {Smith}, \citenamefont {Palizdar}, \citenamefont {Schmidt}, \citenamefont {Bell}, \citenamefont {Coleman},\ and\ \citenamefont {Whatmore}}]{Keeney2020}%
  \BibitemOpen
  \bibfield  {author} {\bibinfo {author} {\bibfnamefont {L.}~\bibnamefont {Keeney}}, \bibinfo {author} {\bibfnamefont {R.~J.}\ \bibnamefont {Smith}}, \bibinfo {author} {\bibfnamefont {M.}~\bibnamefont {Palizdar}}, \bibinfo {author} {\bibfnamefont {M.}~\bibnamefont {Schmidt}}, \bibinfo {author} {\bibfnamefont {A.~J.}\ \bibnamefont {Bell}}, \bibinfo {author} {\bibfnamefont {J.~N.}\ \bibnamefont {Coleman}},\ and\ \bibinfo {author} {\bibfnamefont {R.~W.}\ \bibnamefont {Whatmore}},\ }\href {https://doi.org/10.1002/aelm.201901264} {\bibfield  {journal} {\bibinfo  {journal} {Advanced Electronic Materials}\ }\textbf {\bibinfo {volume} {6}},\ \bibinfo {pages} {1901264} (\bibinfo {year} {2020}{\natexlab{b}})}\BibitemShut {NoStop}%
\bibitem [{\citenamefont {Gradauskaite}\ \emph {et~al.}(2022)\citenamefont {Gradauskaite}, \citenamefont {Hunnestad}, \citenamefont {Meier}, \citenamefont {Meier},\ and\ \citenamefont {Trassin}}]{Gradauskaite2022a}%
  \BibitemOpen
  \bibfield  {author} {\bibinfo {author} {\bibfnamefont {E.}~\bibnamefont {Gradauskaite}}, \bibinfo {author} {\bibfnamefont {K.~A.}\ \bibnamefont {Hunnestad}}, \bibinfo {author} {\bibfnamefont {Q.~N.}\ \bibnamefont {Meier}}, \bibinfo {author} {\bibfnamefont {D.}~\bibnamefont {Meier}},\ and\ \bibinfo {author} {\bibfnamefont {M.}~\bibnamefont {Trassin}},\ }\href {https://doi.org/10.1021/acs.chemmater.2c01178} {\bibfield  {journal} {\bibinfo  {journal} {Chemistry of Materials}\ }\textbf {\bibinfo {volume} {34}},\ \bibinfo {pages} {6468} (\bibinfo {year} {2022})}\BibitemShut {NoStop}%
\bibitem [{\citenamefont {Moore}\ \emph {et~al.}(2022)\citenamefont {Moore}, \citenamefont {O’Connell}, \citenamefont {Griffin}, \citenamefont {Downing}, \citenamefont {Colfer}, \citenamefont {Schmidt}, \citenamefont {Nicolosi}, \citenamefont {Bangert}, \citenamefont {Keeney},\ and\ \citenamefont {Conroy}}]{Moore2022}%
  \BibitemOpen
  \bibfield  {author} {\bibinfo {author} {\bibfnamefont {K.}~\bibnamefont {Moore}}, \bibinfo {author} {\bibfnamefont {E.~N.}\ \bibnamefont {O’Connell}}, \bibinfo {author} {\bibfnamefont {S.~M.}\ \bibnamefont {Griffin}}, \bibinfo {author} {\bibfnamefont {C.}~\bibnamefont {Downing}}, \bibinfo {author} {\bibfnamefont {L.}~\bibnamefont {Colfer}}, \bibinfo {author} {\bibfnamefont {M.}~\bibnamefont {Schmidt}}, \bibinfo {author} {\bibfnamefont {V.}~\bibnamefont {Nicolosi}}, \bibinfo {author} {\bibfnamefont {U.}~\bibnamefont {Bangert}}, \bibinfo {author} {\bibfnamefont {L.}~\bibnamefont {Keeney}},\ and\ \bibinfo {author} {\bibfnamefont {M.}~\bibnamefont {Conroy}},\ }\href {https://doi.org/10.1021/acsami.1c17383} {\bibfield  {journal} {\bibinfo  {journal} {ACS Applied Materials \& Interfaces}\ }\textbf {\bibinfo {volume} {14}},\ \bibinfo {pages} {5525} (\bibinfo {year} {2022})}\BibitemShut {NoStop}%
\bibitem [{\citenamefont {Gradauskaite}\ \emph {et~al.}(2023)\citenamefont {Gradauskaite}, \citenamefont {Meier}, \citenamefont {Gray}, \citenamefont {Sarott}, \citenamefont {Scharsach}, \citenamefont {Campanini}, \citenamefont {Moran}, \citenamefont {Vogel}, \citenamefont {Del Cid-Ledezma}, \citenamefont {Huey}, \citenamefont {Rossell}, \citenamefont {Fiebig},\ and\ \citenamefont {Trassin}}]{gradauskaite_defeating_2023}%
  \BibitemOpen
  \bibfield  {author} {\bibinfo {author} {\bibfnamefont {E.}~\bibnamefont {Gradauskaite}}, \bibinfo {author} {\bibfnamefont {Q.~N.}\ \bibnamefont {Meier}}, \bibinfo {author} {\bibfnamefont {N.}~\bibnamefont {Gray}}, \bibinfo {author} {\bibfnamefont {M.~F.}\ \bibnamefont {Sarott}}, \bibinfo {author} {\bibfnamefont {T.}~\bibnamefont {Scharsach}}, \bibinfo {author} {\bibfnamefont {M.}~\bibnamefont {Campanini}}, \bibinfo {author} {\bibfnamefont {T.}~\bibnamefont {Moran}}, \bibinfo {author} {\bibfnamefont {A.}~\bibnamefont {Vogel}}, \bibinfo {author} {\bibfnamefont {K.}~\bibnamefont {Del Cid-Ledezma}}, \bibinfo {author} {\bibfnamefont {B.~D.}\ \bibnamefont {Huey}}, \bibinfo {author} {\bibfnamefont {M.~D.}\ \bibnamefont {Rossell}}, \bibinfo {author} {\bibfnamefont {M.}~\bibnamefont {Fiebig}},\ and\ \bibinfo {author} {\bibfnamefont {M.}~\bibnamefont {Trassin}},\ }\href {https://doi.org/10.1038/s41563-023-01674-2} {\bibfield  {journal} {\bibinfo  {journal} {Nature Materials}\ }\textbf {\bibinfo {volume} {22}},\
  \bibinfo {pages} {1492} (\bibinfo {year} {2023})}\BibitemShut {NoStop}%
\bibitem [{\citenamefont {Bednorz}\ and\ \citenamefont {Mueller}(1986)}]{bednorz_possible_1986}%
  \BibitemOpen
  \bibfield  {author} {\bibinfo {author} {\bibfnamefont {J.~G.}\ \bibnamefont {Bednorz}}\ and\ \bibinfo {author} {\bibfnamefont {K.~A.}\ \bibnamefont {Mueller}},\ }\href {https://doi.org/10.1007/BF01303701} {\bibfield  {journal} {\bibinfo  {journal} {Zeitschrift fur Physik B Condensed Matter}\ }\textbf {\bibinfo {volume} {64}},\ \bibinfo {pages} {189} (\bibinfo {year} {1986})}\BibitemShut {NoStop}%
\bibitem [{\citenamefont {Maeno}\ \emph {et~al.}(1994)\citenamefont {Maeno}, \citenamefont {Hashimoto}, \citenamefont {Yoshida}, \citenamefont {Nishizaki}, \citenamefont {Fujita}, \citenamefont {Bednorz},\ and\ \citenamefont {Lichtenberg}}]{maeno_superconductivity_1994}%
  \BibitemOpen
  \bibfield  {author} {\bibinfo {author} {\bibfnamefont {Y.}~\bibnamefont {Maeno}}, \bibinfo {author} {\bibfnamefont {H.}~\bibnamefont {Hashimoto}}, \bibinfo {author} {\bibfnamefont {K.}~\bibnamefont {Yoshida}}, \bibinfo {author} {\bibfnamefont {S.}~\bibnamefont {Nishizaki}}, \bibinfo {author} {\bibfnamefont {T.}~\bibnamefont {Fujita}}, \bibinfo {author} {\bibfnamefont {J.~G.}\ \bibnamefont {Bednorz}},\ and\ \bibinfo {author} {\bibfnamefont {F.}~\bibnamefont {Lichtenberg}},\ }\href {https://doi.org/10.1038/372532a0} {\bibfield  {journal} {\bibinfo  {journal} {Nature}\ }\textbf {\bibinfo {volume} {372}},\ \bibinfo {pages} {532} (\bibinfo {year} {1994})}\BibitemShut {NoStop}%
\bibitem [{\citenamefont {A-Paz~de Araujo}\ \emph {et~al.}(1995)\citenamefont {A-Paz~de Araujo}, \citenamefont {Cuchiaro}, \citenamefont {McMillian}, \citenamefont {Scott},\ and\ \citenamefont {Scott}}]{a-paz_de_araujo_fatigue-free_1995}%
  \BibitemOpen
  \bibfield  {author} {\bibinfo {author} {\bibfnamefont {C.}~\bibnamefont {A-Paz~de Araujo}}, \bibinfo {author} {\bibfnamefont {J.~D.}\ \bibnamefont {Cuchiaro}}, \bibinfo {author} {\bibfnamefont {L.~D.}\ \bibnamefont {McMillian}}, \bibinfo {author} {\bibfnamefont {M.~C.}\ \bibnamefont {Scott}},\ and\ \bibinfo {author} {\bibfnamefont {J.~F.}\ \bibnamefont {Scott}},\ }\href {https://doi.org/10.1038/374627a0} {\bibfield  {journal} {\bibinfo  {journal} {Nature}\ }\textbf {\bibinfo {volume} {374}},\ \bibinfo {pages} {627} (\bibinfo {year} {1995})}\BibitemShut {NoStop}%
\bibitem [{\citenamefont {Carpy}\ and\ \citenamefont {Galy}(1974)}]{carpy_systeme_1974}%
  \BibitemOpen
  \bibfield  {author} {\bibinfo {author} {\bibfnamefont {A.}~\bibnamefont {Carpy}}\ and\ \bibinfo {author} {\bibfnamefont {J.}~\bibnamefont {Galy}},\ }\href {https://doi.org/10.3406/bulmi.1974.6938} {\bibfield  {journal} {\bibinfo  {journal} {Bulletin de la Société française de Minéralogie et de Cristallographie}\ }\textbf {\bibinfo {volume} {97}},\ \bibinfo {pages} {484} (\bibinfo {year} {1974})}\BibitemShut {NoStop}%
\bibitem [{\citenamefont {Nanamatsu}\ \emph {et~al.}(1975)\citenamefont {Nanamatsu}, \citenamefont {Kimura},\ and\ \citenamefont {Kawamura}}]{nanamatsu_crystallographic_1975}%
  \BibitemOpen
  \bibfield  {author} {\bibinfo {author} {\bibfnamefont {S.}~\bibnamefont {Nanamatsu}}, \bibinfo {author} {\bibfnamefont {M.}~\bibnamefont {Kimura}},\ and\ \bibinfo {author} {\bibfnamefont {T.}~\bibnamefont {Kawamura}},\ }\href {https://doi.org/10.1143/JPSJ.38.817} {\bibfield  {journal} {\bibinfo  {journal} {Journal of the Physical Society of Japan}\ }\textbf {\bibinfo {volume} {38}},\ \bibinfo {pages} {817} (\bibinfo {year} {1975})}\BibitemShut {NoStop}%
\bibitem [{\citenamefont {Isupov}(1999)}]{isupovCrystalChemicalAspects1999}%
  \BibitemOpen
  \bibfield  {author} {\bibinfo {author} {\bibfnamefont {V.~A.}\ \bibnamefont {Isupov}},\ }\href {https://doi.org/10.1080/00150199908007997} {\bibfield  {journal} {\bibinfo  {journal} {Ferroelectrics}\ }\textbf {\bibinfo {volume} {220}},\ \bibinfo {pages} {79} (\bibinfo {year} {1999})}\BibitemShut {NoStop}%
\bibitem [{\citenamefont {Lichtenberg}(2001)}]{lichtenbergSynthesisPerovskiterelatedLayered2001}%
  \BibitemOpen
  \bibfield  {author} {\bibinfo {author} {\bibfnamefont {F.}~\bibnamefont {Lichtenberg}},\ }\href {https://doi.org/10.1016/S0079-6786(01)00002-4} {\bibfield  {journal} {\bibinfo  {journal} {Progress in Solid State Chemistry}\ }\textbf {\bibinfo {volume} {29}},\ \bibinfo {pages} {1} (\bibinfo {year} {2001})}\BibitemShut {NoStop}%
\bibitem [{\citenamefont {Lichtenberg}\ \emph {et~al.}(2008)\citenamefont {Lichtenberg}, \citenamefont {Herrnberger},\ and\ \citenamefont {Wiedenmann}}]{lichtenbergSynthesisStructuralMagnetic2008}%
  \BibitemOpen
  \bibfield  {author} {\bibinfo {author} {\bibfnamefont {F.}~\bibnamefont {Lichtenberg}}, \bibinfo {author} {\bibfnamefont {A.}~\bibnamefont {Herrnberger}},\ and\ \bibinfo {author} {\bibfnamefont {K.}~\bibnamefont {Wiedenmann}},\ }\href {https://doi.org/10.1016/j.progsolidstchem.2008.10.001} {\bibfield  {journal} {\bibinfo  {journal} {Progress in Solid State Chemistry}\ }\textbf {\bibinfo {volume} {36}},\ \bibinfo {pages} {253} (\bibinfo {year} {2008})}\BibitemShut {NoStop}%
\bibitem [{\citenamefont {Núñez~Valdez}\ and\ \citenamefont {Spaldin}(2019)}]{nunez_valdez_origin_2019}%
  \BibitemOpen
  \bibfield  {author} {\bibinfo {author} {\bibfnamefont {M.}~\bibnamefont {Núñez~Valdez}}\ and\ \bibinfo {author} {\bibfnamefont {N.~A.}\ \bibnamefont {Spaldin}},\ }\href {https://doi.org/10.1016/j.poly.2019.07.018} {\bibfield  {journal} {\bibinfo  {journal} {Polyhedron}\ }\textbf {\bibinfo {volume} {171}},\ \bibinfo {pages} {181} (\bibinfo {year} {2019})}\BibitemShut {NoStop}%
\bibitem [{\citenamefont {López-Pérez}\ and\ \citenamefont {Íñiguez}(2011)}]{lopez-perezInitioStudyProper2011}%
  \BibitemOpen
  \bibfield  {author} {\bibinfo {author} {\bibfnamefont {J.}~\bibnamefont {López-Pérez}}\ and\ \bibinfo {author} {\bibfnamefont {J.}~\bibnamefont {Íñiguez}},\ }\href {https://doi.org/10.1103/PhysRevB.84.075121} {\bibfield  {journal} {\bibinfo  {journal} {Physical Review B}\ }\textbf {\bibinfo {volume} {84}},\ \bibinfo {pages} {075121} (\bibinfo {year} {2011})}\BibitemShut {NoStop}%
\bibitem [{\citenamefont {Daniels}\ \emph {et~al.}(2002)\citenamefont {Daniels}, \citenamefont {Tamazyan}, \citenamefont {Kuntscher}, \citenamefont {Dressel}, \citenamefont {Lichtenberg},\ and\ \citenamefont {Van~Smaalen}}]{daniels_incommensurate_2002}%
  \BibitemOpen
  \bibfield  {author} {\bibinfo {author} {\bibfnamefont {P.}~\bibnamefont {Daniels}}, \bibinfo {author} {\bibfnamefont {R.}~\bibnamefont {Tamazyan}}, \bibinfo {author} {\bibfnamefont {C.~A.}\ \bibnamefont {Kuntscher}}, \bibinfo {author} {\bibfnamefont {M.}~\bibnamefont {Dressel}}, \bibinfo {author} {\bibfnamefont {F.}~\bibnamefont {Lichtenberg}},\ and\ \bibinfo {author} {\bibfnamefont {S.}~\bibnamefont {Van~Smaalen}},\ }\href {https://doi.org/10.1107/S010876810201741X} {\bibfield  {journal} {\bibinfo  {journal} {Acta Crystallographica Section B Structural Science}\ }\textbf {\bibinfo {volume} {58}},\ \bibinfo {pages} {970} (\bibinfo {year} {2002})}\BibitemShut {NoStop}%
\bibitem [{\citenamefont {Howieson}\ \emph {et~al.}(2020)\citenamefont {Howieson}, \citenamefont {Wu}, \citenamefont {Gibbs}, \citenamefont {Zhou}, \citenamefont {Scott},\ and\ \citenamefont {Morrison}}]{howiesonIncommensurateCommensurateTransition2020}%
  \BibitemOpen
  \bibfield  {author} {\bibinfo {author} {\bibfnamefont {G.~W.}\ \bibnamefont {Howieson}}, \bibinfo {author} {\bibfnamefont {S.}~\bibnamefont {Wu}}, \bibinfo {author} {\bibfnamefont {A.~S.}\ \bibnamefont {Gibbs}}, \bibinfo {author} {\bibfnamefont {W.}~\bibnamefont {Zhou}}, \bibinfo {author} {\bibfnamefont {J.~F.}\ \bibnamefont {Scott}},\ and\ \bibinfo {author} {\bibfnamefont {F.~D.}\ \bibnamefont {Morrison}},\ }\href {https://doi.org/10.1002/adfm.202004667} {\bibfield  {journal} {\bibinfo  {journal} {Advanced Functional Materials}\ }\textbf {\bibinfo {volume} {30}},\ \bibinfo {pages} {2004667} (\bibinfo {year} {2020})}\BibitemShut {NoStop}%
\bibitem [{\citenamefont {Ederer}\ and\ \citenamefont {Spaldin}(2006)}]{ederer_electric-field-switchable_2006}%
  \BibitemOpen
  \bibfield  {author} {\bibinfo {author} {\bibfnamefont {C.}~\bibnamefont {Ederer}}\ and\ \bibinfo {author} {\bibfnamefont {N.~A.}\ \bibnamefont {Spaldin}},\ }\href {https://doi.org/10.1103/PhysRevB.74.020401} {\bibfield  {journal} {\bibinfo  {journal} {Physical Review B}\ }\textbf {\bibinfo {volume} {74}},\ \bibinfo {pages} {020401} (\bibinfo {year} {2006})}\BibitemShut {NoStop}%
\bibitem [{\citenamefont {Kuntscher}\ \emph {et~al.}(2004)\citenamefont {Kuntscher}, \citenamefont {Schuppler}, \citenamefont {Haas}, \citenamefont {Gorshunov}, \citenamefont {Dressel}, \citenamefont {Grioni},\ and\ \citenamefont {Lichtenberg}}]{Kuntscher2004}%
  \BibitemOpen
  \bibfield  {author} {\bibinfo {author} {\bibfnamefont {C.~A.}\ \bibnamefont {Kuntscher}}, \bibinfo {author} {\bibfnamefont {S.}~\bibnamefont {Schuppler}}, \bibinfo {author} {\bibfnamefont {P.}~\bibnamefont {Haas}}, \bibinfo {author} {\bibfnamefont {B.}~\bibnamefont {Gorshunov}}, \bibinfo {author} {\bibfnamefont {M.}~\bibnamefont {Dressel}}, \bibinfo {author} {\bibfnamefont {M.}~\bibnamefont {Grioni}},\ and\ \bibinfo {author} {\bibfnamefont {F.}~\bibnamefont {Lichtenberg}},\ }\href {https://doi.org/10.1103/PhysRevB.70.245123} {\bibfield  {journal} {\bibinfo  {journal} {Physical Review B - Condensed Matter and Materials Physics}\ }\textbf {\bibinfo {volume} {70}},\ \bibinfo {pages} {1} (\bibinfo {year} {2004})}\BibitemShut {NoStop}%
\bibitem [{\citenamefont {Nanamatsu}\ \emph {et~al.}(1974)\citenamefont {Nanamatsu}, \citenamefont {Kimura}, \citenamefont {Doi}, \citenamefont {Matsushita},\ and\ \citenamefont {Yamada}}]{nanamatsuNewFerroelectricLa2Ti2o71974}%
  \BibitemOpen
  \bibfield  {author} {\bibinfo {author} {\bibfnamefont {S.}~\bibnamefont {Nanamatsu}}, \bibinfo {author} {\bibfnamefont {M.}~\bibnamefont {Kimura}}, \bibinfo {author} {\bibfnamefont {K.}~\bibnamefont {Doi}}, \bibinfo {author} {\bibfnamefont {S.}~\bibnamefont {Matsushita}},\ and\ \bibinfo {author} {\bibfnamefont {N.}~\bibnamefont {Yamada}},\ }\href {https://doi.org/10.1080/00150197408234143} {\bibfield  {journal} {\bibinfo  {journal} {Ferroelectrics}\ }\textbf {\bibinfo {volume} {8}},\ \bibinfo {pages} {511} (\bibinfo {year} {1974})}\BibitemShut {NoStop}%
\bibitem [{\citenamefont {Schmalle}\ \emph {et~al.}(1993)\citenamefont {Schmalle}, \citenamefont {Williams}, \citenamefont {Reller}, \citenamefont {Linden},\ and\ \citenamefont {Bednorz}}]{schmalle_twin_1993}%
  \BibitemOpen
  \bibfield  {author} {\bibinfo {author} {\bibfnamefont {H.~W.}\ \bibnamefont {Schmalle}}, \bibinfo {author} {\bibfnamefont {T.}~\bibnamefont {Williams}}, \bibinfo {author} {\bibfnamefont {A.}~\bibnamefont {Reller}}, \bibinfo {author} {\bibfnamefont {A.}~\bibnamefont {Linden}},\ and\ \bibinfo {author} {\bibfnamefont {J.~G.}\ \bibnamefont {Bednorz}},\ }\href {https://doi.org/10.1107/S010876819200987X} {\bibfield  {journal} {\bibinfo  {journal} {Acta Crystallographica Section B Structural Science}\ }\textbf {\bibinfo {volume} {49}},\ \bibinfo {pages} {235} (\bibinfo {year} {1993})}\BibitemShut {NoStop}%
\bibitem [{\citenamefont {Havelia}\ \emph {et~al.}(2009{\natexlab{a}})\citenamefont {Havelia}, \citenamefont {Wang}, \citenamefont {Balasubramaniam},\ and\ \citenamefont {Salvador}}]{havelia2009}%
  \BibitemOpen
  \bibfield  {author} {\bibinfo {author} {\bibfnamefont {S.}~\bibnamefont {Havelia}}, \bibinfo {author} {\bibfnamefont {S.}~\bibnamefont {Wang}}, \bibinfo {author} {\bibfnamefont {K.~R.}\ \bibnamefont {Balasubramaniam}},\ and\ \bibinfo {author} {\bibfnamefont {P.~A.}\ \bibnamefont {Salvador}},\ }\href {https://doi.org/10.1021/cg900556d} {\bibfield  {journal} {\bibinfo  {journal} {Crystal Growth \& Design}\ }\textbf {\bibinfo {volume} {9}},\ \bibinfo {pages} {4546} (\bibinfo {year} {2009}{\natexlab{a}})}\BibitemShut {NoStop}%
\bibitem [{\citenamefont {Ohtomo}\ \emph {et~al.}(2002)\citenamefont {Ohtomo}, \citenamefont {Muller}, \citenamefont {Grazul},\ and\ \citenamefont {Hwang}}]{ohtomoEpitaxialGrowthElectronic2002}%
  \BibitemOpen
  \bibfield  {author} {\bibinfo {author} {\bibfnamefont {A.}~\bibnamefont {Ohtomo}}, \bibinfo {author} {\bibfnamefont {D.~A.}\ \bibnamefont {Muller}}, \bibinfo {author} {\bibfnamefont {J.~L.}\ \bibnamefont {Grazul}},\ and\ \bibinfo {author} {\bibfnamefont {H.~Y.}\ \bibnamefont {Hwang}},\ }\href {https://doi.org/10.1063/1.1481767} {\bibfield  {journal} {\bibinfo  {journal} {Applied Physics Letters}\ }\textbf {\bibinfo {volume} {80}},\ \bibinfo {pages} {3922} (\bibinfo {year} {2002})}\BibitemShut {NoStop}%
\bibitem [{\citenamefont {Havelia}\ \emph {et~al.}(2008)\citenamefont {Havelia}, \citenamefont {Balasubramaniam}, \citenamefont {Spurgeon}, \citenamefont {Cormack},\ and\ \citenamefont {Salvador}}]{haveliaGrowthLa2Ti2o7LaTiO32008}%
  \BibitemOpen
  \bibfield  {author} {\bibinfo {author} {\bibfnamefont {S.}~\bibnamefont {Havelia}}, \bibinfo {author} {\bibfnamefont {K.}~\bibnamefont {Balasubramaniam}}, \bibinfo {author} {\bibfnamefont {S.}~\bibnamefont {Spurgeon}}, \bibinfo {author} {\bibfnamefont {F.}~\bibnamefont {Cormack}},\ and\ \bibinfo {author} {\bibfnamefont {P.}~\bibnamefont {Salvador}},\ }\href {https://doi.org/10.1016/j.jcrysgro.2007.12.006} {\bibfield  {journal} {\bibinfo  {journal} {Journal of Crystal Growth}\ }\textbf {\bibinfo {volume} {310}},\ \bibinfo {pages} {1985} (\bibinfo {year} {2008})}\BibitemShut {NoStop}%
\bibitem [{\citenamefont {Havelia}\ \emph {et~al.}(2009{\natexlab{b}})\citenamefont {Havelia}, \citenamefont {Wang}, \citenamefont {Balasubramaniam},\ and\ \citenamefont {Salvador}}]{havelia2009a}%
  \BibitemOpen
  \bibfield  {author} {\bibinfo {author} {\bibfnamefont {S.}~\bibnamefont {Havelia}}, \bibinfo {author} {\bibfnamefont {S.}~\bibnamefont {Wang}}, \bibinfo {author} {\bibfnamefont {K.}~\bibnamefont {Balasubramaniam}},\ and\ \bibinfo {author} {\bibfnamefont {P.}~\bibnamefont {Salvador}},\ }\href {https://doi.org/10.1016/j.jssc.2009.02.025} {\bibfield  {journal} {\bibinfo  {journal} {Journal of Solid State Chemistry}\ }\textbf {\bibinfo {volume} {182}},\ \bibinfo {pages} {1603} (\bibinfo {year} {2009}{\natexlab{b}})}\BibitemShut {NoStop}%
\bibitem [{\citenamefont {Kaspar}\ \emph {et~al.}(2018)\citenamefont {Kaspar}, \citenamefont {Hong}, \citenamefont {Bowden}, \citenamefont {Varga}, \citenamefont {Yan}, \citenamefont {Wang}, \citenamefont {Spurgeon}, \citenamefont {Comes}, \citenamefont {Ramuhalli},\ and\ \citenamefont {Henager}}]{kasparTuningPiezoelectricProperties2018}%
  \BibitemOpen
  \bibfield  {author} {\bibinfo {author} {\bibfnamefont {T.~C.}\ \bibnamefont {Kaspar}}, \bibinfo {author} {\bibfnamefont {S.}~\bibnamefont {Hong}}, \bibinfo {author} {\bibfnamefont {M.~E.}\ \bibnamefont {Bowden}}, \bibinfo {author} {\bibfnamefont {T.}~\bibnamefont {Varga}}, \bibinfo {author} {\bibfnamefont {P.}~\bibnamefont {Yan}}, \bibinfo {author} {\bibfnamefont {C.}~\bibnamefont {Wang}}, \bibinfo {author} {\bibfnamefont {S.~R.}\ \bibnamefont {Spurgeon}}, \bibinfo {author} {\bibfnamefont {R.~B.}\ \bibnamefont {Comes}}, \bibinfo {author} {\bibfnamefont {P.}~\bibnamefont {Ramuhalli}},\ and\ \bibinfo {author} {\bibfnamefont {C.~H.}\ \bibnamefont {Henager}},\ }\href {https://doi.org/10.1038/s41598-018-21009-5} {\bibfield  {journal} {\bibinfo  {journal} {Scientific Reports}\ }\textbf {\bibinfo {volume} {8}},\ \bibinfo {pages} {3037} (\bibinfo {year} {2018})}\BibitemShut {NoStop}%
\bibitem [{\citenamefont {Bayart}\ \emph {et~al.}(2013)\citenamefont {Bayart}, \citenamefont {Saitzek}, \citenamefont {Chambrier}, \citenamefont {Shao}, \citenamefont {Ferri}, \citenamefont {Huvé}, \citenamefont {Pouhet}, \citenamefont {Tebano}, \citenamefont {Roussel},\ and\ \citenamefont {Desfeux}}]{bayart2013}%
  \BibitemOpen
  \bibfield  {author} {\bibinfo {author} {\bibfnamefont {A.}~\bibnamefont {Bayart}}, \bibinfo {author} {\bibfnamefont {S.}~\bibnamefont {Saitzek}}, \bibinfo {author} {\bibfnamefont {M.-H.}\ \bibnamefont {Chambrier}}, \bibinfo {author} {\bibfnamefont {Z.}~\bibnamefont {Shao}}, \bibinfo {author} {\bibfnamefont {A.}~\bibnamefont {Ferri}}, \bibinfo {author} {\bibfnamefont {M.}~\bibnamefont {Huvé}}, \bibinfo {author} {\bibfnamefont {R.}~\bibnamefont {Pouhet}}, \bibinfo {author} {\bibfnamefont {A.}~\bibnamefont {Tebano}}, \bibinfo {author} {\bibfnamefont {P.}~\bibnamefont {Roussel}},\ and\ \bibinfo {author} {\bibfnamefont {R.}~\bibnamefont {Desfeux}},\ }\href {https://doi.org/10.1039/c3ce40256h} {\bibfield  {journal} {\bibinfo  {journal} {CrystEngComm}\ }\textbf {\bibinfo {volume} {15}},\ \bibinfo {pages} {4341} (\bibinfo {year} {2013})}\BibitemShut {NoStop}%
\bibitem [{\citenamefont {Ishizawa}\ \emph {et~al.}(1982)\citenamefont {Ishizawa}, \citenamefont {Marumo}, \citenamefont {Iwai}, \citenamefont {Kimura},\ and\ \citenamefont {Kawamura}}]{ishizawa_compounds_1982}%
  \BibitemOpen
  \bibfield  {author} {\bibinfo {author} {\bibfnamefont {N.}~\bibnamefont {Ishizawa}}, \bibinfo {author} {\bibfnamefont {F.}~\bibnamefont {Marumo}}, \bibinfo {author} {\bibfnamefont {S.}~\bibnamefont {Iwai}}, \bibinfo {author} {\bibfnamefont {M.}~\bibnamefont {Kimura}},\ and\ \bibinfo {author} {\bibfnamefont {T.}~\bibnamefont {Kawamura}},\ }\href {https://doi.org/10.1107/S0567740882002994} {\bibfield  {journal} {\bibinfo  {journal} {Acta Crystallographica Section B: Structural Crystallography and Crystal Chemistry}\ }\textbf {\bibinfo {volume} {38}},\ \bibinfo {pages} {368} (\bibinfo {year} {1982})}\BibitemShut {NoStop}%
\bibitem [{\citenamefont {Ishizawa}\ \emph {et~al.}(2019)\citenamefont {Ishizawa}, \citenamefont {Ninomiya},\ and\ \citenamefont {Wang}}]{ishizawaStructuralEvolutionTi2019}%
  \BibitemOpen
  \bibfield  {author} {\bibinfo {author} {\bibfnamefont {N.}~\bibnamefont {Ishizawa}}, \bibinfo {author} {\bibfnamefont {K.}~\bibnamefont {Ninomiya}},\ and\ \bibinfo {author} {\bibfnamefont {J.}~\bibnamefont {Wang}},\ }\href {https://doi.org/10.1107/S2052520619002105} {\bibfield  {journal} {\bibinfo  {journal} {Acta Crystallographica Section B Structural Science, Crystal Engineering and Materials}\ }\textbf {\bibinfo {volume} {75}},\ \bibinfo {pages} {257} (\bibinfo {year} {2019})}\BibitemShut {NoStop}%
\bibitem [{\citenamefont {Howieson}\ \emph {et~al.}(2021)\citenamefont {Howieson}, \citenamefont {Mishra}, \citenamefont {Gibbs}, \citenamefont {Katiyar}, \citenamefont {Scott}, \citenamefont {Morrison},\ and\ \citenamefont {Carpenter}}]{Howieson2021}%
  \BibitemOpen
  \bibfield  {author} {\bibinfo {author} {\bibfnamefont {G.~W.}\ \bibnamefont {Howieson}}, \bibinfo {author} {\bibfnamefont {K.~K.}\ \bibnamefont {Mishra}}, \bibinfo {author} {\bibfnamefont {A.~S.}\ \bibnamefont {Gibbs}}, \bibinfo {author} {\bibfnamefont {R.~S.}\ \bibnamefont {Katiyar}}, \bibinfo {author} {\bibfnamefont {J.~F.}\ \bibnamefont {Scott}}, \bibinfo {author} {\bibfnamefont {F.~D.}\ \bibnamefont {Morrison}},\ and\ \bibinfo {author} {\bibfnamefont {M.}~\bibnamefont {Carpenter}},\ }\href {https://doi.org/10.1103/PhysRevB.103.014119} {\bibfield  {journal} {\bibinfo  {journal} {Phys. Rev. B}\ }\textbf {\bibinfo {volume} {103}},\ \bibinfo {pages} {014119} (\bibinfo {year} {2021})}\BibitemShut {NoStop}%
\bibitem [{\citenamefont {Nezu}\ \emph {et~al.}(2017)\citenamefont {Nezu}, \citenamefont {Zhang}, \citenamefont {Chen}, \citenamefont {Ikuhara},\ and\ \citenamefont {Ohta}}]{nezu_solid-phase_2017}%
  \BibitemOpen
  \bibfield  {author} {\bibinfo {author} {\bibfnamefont {Y.}~\bibnamefont {Nezu}}, \bibinfo {author} {\bibfnamefont {Y.-Q.}\ \bibnamefont {Zhang}}, \bibinfo {author} {\bibfnamefont {C.}~\bibnamefont {Chen}}, \bibinfo {author} {\bibfnamefont {Y.}~\bibnamefont {Ikuhara}},\ and\ \bibinfo {author} {\bibfnamefont {H.}~\bibnamefont {Ohta}},\ }\href {https://doi.org/10.1063/1.4997813} {\bibfield  {journal} {\bibinfo  {journal} {Journal of Applied Physics}\ }\textbf {\bibinfo {volume} {122}},\ \bibinfo {pages} {135305} (\bibinfo {year} {2017})}\BibitemShut {NoStop}%
\bibitem [{\citenamefont {Yao}\ \emph {et~al.}(2020)\citenamefont {Yao}, \citenamefont {Jiang}, \citenamefont {Chen}, \citenamefont {Yan}, \citenamefont {Tao}, \citenamefont {Yang}, \citenamefont {Li}, \citenamefont {Sugo}, \citenamefont {Ohta}, \citenamefont {Ye}, \citenamefont {Ikuhara},\ and\ \citenamefont {Ma}}]{Yao2020}%
  \BibitemOpen
  \bibfield  {author} {\bibinfo {author} {\bibfnamefont {T.}~\bibnamefont {Yao}}, \bibinfo {author} {\bibfnamefont {Y.}~\bibnamefont {Jiang}}, \bibinfo {author} {\bibfnamefont {C.}~\bibnamefont {Chen}}, \bibinfo {author} {\bibfnamefont {X.}~\bibnamefont {Yan}}, \bibinfo {author} {\bibfnamefont {A.}~\bibnamefont {Tao}}, \bibinfo {author} {\bibfnamefont {L.}~\bibnamefont {Yang}}, \bibinfo {author} {\bibfnamefont {C.}~\bibnamefont {Li}}, \bibinfo {author} {\bibfnamefont {K.}~\bibnamefont {Sugo}}, \bibinfo {author} {\bibfnamefont {H.}~\bibnamefont {Ohta}}, \bibinfo {author} {\bibfnamefont {H.}~\bibnamefont {Ye}}, \bibinfo {author} {\bibfnamefont {Y.}~\bibnamefont {Ikuhara}},\ and\ \bibinfo {author} {\bibfnamefont {X.}~\bibnamefont {Ma}},\ }\href {https://doi.org/10.1021/acs.nanolett.9b04210} {\bibfield  {journal} {\bibinfo  {journal} {Nano Letters}\ }\textbf {\bibinfo {volume} {20}},\ \bibinfo {pages} {1047} (\bibinfo {year} {2020})}\BibitemShut {NoStop}%
\bibitem [{\citenamefont {Qiao}\ \emph {et~al.}(2022)\citenamefont {Qiao}, \citenamefont {Jiang}, \citenamefont {Yao}, \citenamefont {Tao}, \citenamefont {Yan}, \citenamefont {Gao}, \citenamefont {Li}, \citenamefont {Ohta}, \citenamefont {Chen}, \citenamefont {Ma},\ and\ \citenamefont {Ye}}]{qiaoMicrostructureElectricalProperties2022}%
  \BibitemOpen
  \bibfield  {author} {\bibinfo {author} {\bibfnamefont {B.}~\bibnamefont {Qiao}}, \bibinfo {author} {\bibfnamefont {Y.}~\bibnamefont {Jiang}}, \bibinfo {author} {\bibfnamefont {T.}~\bibnamefont {Yao}}, \bibinfo {author} {\bibfnamefont {A.}~\bibnamefont {Tao}}, \bibinfo {author} {\bibfnamefont {X.}~\bibnamefont {Yan}}, \bibinfo {author} {\bibfnamefont {C.}~\bibnamefont {Gao}}, \bibinfo {author} {\bibfnamefont {X.}~\bibnamefont {Li}}, \bibinfo {author} {\bibfnamefont {H.}~\bibnamefont {Ohta}}, \bibinfo {author} {\bibfnamefont {C.}~\bibnamefont {Chen}}, \bibinfo {author} {\bibfnamefont {X.-L.}\ \bibnamefont {Ma}},\ and\ \bibinfo {author} {\bibfnamefont {H.}~\bibnamefont {Ye}},\ }\href {https://doi.org/10.1016/j.apsusc.2021.151599} {\bibfield  {journal} {\bibinfo  {journal} {Applied Surface Science}\ }\textbf {\bibinfo {volume} {587}},\ \bibinfo {pages} {151599} (\bibinfo {year} {2022})}\BibitemShut {NoStop}%
\bibitem [{\citenamefont {Shao}\ \emph {et~al.}(2011)\citenamefont {Shao}, \citenamefont {Saitzek}, \citenamefont {Roussel}, \citenamefont {Ferri}, \citenamefont {Bruyer}, \citenamefont {Sayede}, \citenamefont {Rguiti}, \citenamefont {Mentre},\ and\ \citenamefont {Desfeux}}]{shao_microstructure_2011}%
  \BibitemOpen
  \bibfield  {author} {\bibinfo {author} {\bibfnamefont {Z.}~\bibnamefont {Shao}}, \bibinfo {author} {\bibfnamefont {S.}~\bibnamefont {Saitzek}}, \bibinfo {author} {\bibfnamefont {P.}~\bibnamefont {Roussel}}, \bibinfo {author} {\bibfnamefont {A.}~\bibnamefont {Ferri}}, \bibinfo {author} {\bibfnamefont {E.}~\bibnamefont {Bruyer}}, \bibinfo {author} {\bibfnamefont {A.}~\bibnamefont {Sayede}}, \bibinfo {author} {\bibfnamefont {M.}~\bibnamefont {Rguiti}}, \bibinfo {author} {\bibfnamefont {O.}~\bibnamefont {Mentre}},\ and\ \bibinfo {author} {\bibfnamefont {R.}~\bibnamefont {Desfeux}},\ }\href {https://doi.org/10.1002/adem.201100105} {\bibfield  {journal} {\bibinfo  {journal} {Advanced Engineering Materials}\ }\textbf {\bibinfo {volume} {13}},\ \bibinfo {pages} {961} (\bibinfo {year} {2011})}\BibitemShut {NoStop}%
\bibitem [{\citenamefont {Zurbuchen}\ \emph {et~al.}(2007)\citenamefont {Zurbuchen}, \citenamefont {Tian}, \citenamefont {Pan}, \citenamefont {Fong}, \citenamefont {Streiffer}, \citenamefont {Hawley}, \citenamefont {Lettieri}, \citenamefont {Jia}, \citenamefont {Asayama}, \citenamefont {Fulk}, \citenamefont {Comstock}, \citenamefont {Knapp}, \citenamefont {Carim},\ and\ \citenamefont {Schlom}}]{Zurbuchen2007}%
  \BibitemOpen
  \bibfield  {author} {\bibinfo {author} {\bibfnamefont {M.~A.}\ \bibnamefont {Zurbuchen}}, \bibinfo {author} {\bibfnamefont {W.}~\bibnamefont {Tian}}, \bibinfo {author} {\bibfnamefont {X.~Q.}\ \bibnamefont {Pan}}, \bibinfo {author} {\bibfnamefont {D.}~\bibnamefont {Fong}}, \bibinfo {author} {\bibfnamefont {S.~K.}\ \bibnamefont {Streiffer}}, \bibinfo {author} {\bibfnamefont {M.~E.}\ \bibnamefont {Hawley}}, \bibinfo {author} {\bibfnamefont {J.}~\bibnamefont {Lettieri}}, \bibinfo {author} {\bibfnamefont {Y.}~\bibnamefont {Jia}}, \bibinfo {author} {\bibfnamefont {G.}~\bibnamefont {Asayama}}, \bibinfo {author} {\bibfnamefont {S.~J.}\ \bibnamefont {Fulk}}, \bibinfo {author} {\bibfnamefont {D.~J.}\ \bibnamefont {Comstock}}, \bibinfo {author} {\bibfnamefont {S.}~\bibnamefont {Knapp}}, \bibinfo {author} {\bibfnamefont {A.~H.}\ \bibnamefont {Carim}},\ and\ \bibinfo {author} {\bibfnamefont {D.~G.}\ \bibnamefont {Schlom}},\ }\href {https://doi.org/10.1557/jmr.2007.0198} {\bibfield  {journal} {\bibinfo  {journal}
  {Journal of Materials Research}\ }\textbf {\bibinfo {volume} {22}},\ \bibinfo {pages} {1439} (\bibinfo {year} {2007})}\BibitemShut {NoStop}%
\bibitem [{\citenamefont {Keeney}\ \emph {et~al.}(2023)\citenamefont {Keeney}, \citenamefont {Colfer}, \citenamefont {Dutta}, \citenamefont {Schmidt},\ and\ \citenamefont {Wei}}]{keeney_what_2023}%
  \BibitemOpen
  \bibfield  {author} {\bibinfo {author} {\bibfnamefont {L.}~\bibnamefont {Keeney}}, \bibinfo {author} {\bibfnamefont {L.}~\bibnamefont {Colfer}}, \bibinfo {author} {\bibfnamefont {D.}~\bibnamefont {Dutta}}, \bibinfo {author} {\bibfnamefont {M.}~\bibnamefont {Schmidt}},\ and\ \bibinfo {author} {\bibfnamefont {G.}~\bibnamefont {Wei}},\ }\bibfield  {journal} {\bibinfo  {journal} {Microstructures}\ }\textbf {\bibinfo {volume} {3}},\ \href {https://doi.org/10.20517/microstructures.2023.41} {10.20517/microstructures.2023.41} (\bibinfo {year} {2023})\BibitemShut {NoStop}%
\bibitem [{\citenamefont {Biswas}\ \emph {et~al.}(2017)\citenamefont {Biswas}, \citenamefont {Yang}, \citenamefont {Ramesh},\ and\ \citenamefont {Jeong}}]{biswas_atomically_2017}%
  \BibitemOpen
  \bibfield  {author} {\bibinfo {author} {\bibfnamefont {A.}~\bibnamefont {Biswas}}, \bibinfo {author} {\bibfnamefont {C.-H.}\ \bibnamefont {Yang}}, \bibinfo {author} {\bibfnamefont {R.}~\bibnamefont {Ramesh}},\ and\ \bibinfo {author} {\bibfnamefont {Y.~H.}\ \bibnamefont {Jeong}},\ }\href {https://doi.org/10.1016/j.progsurf.2017.05.001} {\bibfield  {journal} {\bibinfo  {journal} {Progress in Surface Science}\ }\textbf {\bibinfo {volume} {92}},\ \bibinfo {pages} {117} (\bibinfo {year} {2017})}\BibitemShut {NoStop}%
\bibitem [{\citenamefont {Kresse}\ and\ \citenamefont {Hafner}(1993)}]{kresseInitioMolecularDynamics1993}%
  \BibitemOpen
  \bibfield  {author} {\bibinfo {author} {\bibfnamefont {G.}~\bibnamefont {Kresse}}\ and\ \bibinfo {author} {\bibfnamefont {J.}~\bibnamefont {Hafner}},\ }\href {https://doi.org/10.1103/PhysRevB.47.558} {\bibfield  {journal} {\bibinfo  {journal} {Physical Review B}\ }\textbf {\bibinfo {volume} {47}},\ \bibinfo {pages} {558} (\bibinfo {year} {1993})}\BibitemShut {NoStop}%
\bibitem [{\citenamefont {Kresse}\ and\ \citenamefont {Furthm{\"u}ller}(1996)}]{kresseEfficiencyInitioTotal1996}%
  \BibitemOpen
  \bibfield  {author} {\bibinfo {author} {\bibfnamefont {G.}~\bibnamefont {Kresse}}\ and\ \bibinfo {author} {\bibfnamefont {J.}~\bibnamefont {Furthm{\"u}ller}},\ }\href {https://doi.org/10.1016/0927-0256(96)00008-0} {\bibfield  {journal} {\bibinfo  {journal} {Computational Materials Science}\ }\textbf {\bibinfo {volume} {6}},\ \bibinfo {pages} {15} (\bibinfo {year} {1996})}\BibitemShut {NoStop}%
\bibitem [{\citenamefont {Orobengoa}\ \emph {et~al.}(2009)\citenamefont {Orobengoa}, \citenamefont {Capillas}, \citenamefont {Aroyo},\ and\ \citenamefont {{Perez-Mato}}}]{orobengoaAMPLIMODESSymmetrymodeAnalysis2009}%
  \BibitemOpen
  \bibfield  {author} {\bibinfo {author} {\bibfnamefont {D.}~\bibnamefont {Orobengoa}}, \bibinfo {author} {\bibfnamefont {C.}~\bibnamefont {Capillas}}, \bibinfo {author} {\bibfnamefont {M.~I.}\ \bibnamefont {Aroyo}},\ and\ \bibinfo {author} {\bibfnamefont {J.~M.}\ \bibnamefont {{Perez-Mato}}},\ }\href {https://doi.org/10.1107/S0021889809028064} {\bibfield  {journal} {\bibinfo  {journal} {J. Appl. Crystallogr.}\ }\textbf {\bibinfo {volume} {42}},\ \bibinfo {pages} {820} (\bibinfo {year} {2009})}\BibitemShut {NoStop}%
\end{thebibliography}%

\appendix
\newpage
\section*{Supporting Information}

\renewcommand{\thefigure}{S\arabic{figure}}
\renewcommand{\thetable}{S\arabic{table}}
\setcounter{figure}{1} 
\setcounter{table}{0}  

\subsection*{Supporting Note S1: Epitaxial matching between La\textsubscript{2}Ti\textsubscript{2}O\textsubscript{7} (LTO) of the layered Carpy-Galy phase and SrTiO\textsubscript{3} (STO), DyScO\textsubscript{3} (DSO), and (LaAlO\textsubscript{3})\textsubscript{0.3}(Sr\textsubscript{2}TaAlO\textsubscript{6})\textsubscript{0.7} (LSAT) substrates}
\begin{table}[htb!]
    \centering
    \setlength{\arrayrulewidth}{0.5mm} 
    \renewcommand{\arraystretch}{1.35} 
    \begin{adjustbox}{max width=0.8\textwidth} 
    \begin{tabular}{|l|c|c|c|c|c|c|}
        \hline
        \rowcolor[gray]{0.9} \textbf{Compounds}  & \multicolumn{4}{c|}{\textbf{Lattice parameters (\AA)}} & \multicolumn{2}{c|}{\textbf{LTO Lattice Mismatch}} \\ \hline
        \rowcolor[gray]{0.9} ~  & In-plane 1  & Value  & In-plane 2 (P-axis)  & Value  & along \textit{a} (\%)  & along \textit{b} (\%)  \\ \hline
        La\textsubscript{2}Ti\textsubscript{2}O\textsubscript{7} (LTO)  & \textit{a}  & 7.800  & \textit{b}  & 5.546  & -  & -  \\ \hline
        SrTiO\textsubscript{3} (STO)  & \textit{c*2}  & 7.810  & \(\sqrt{a^2+b^2}\)  & 5.522  & 0.13  & -0.43  \\ \hline
        DyScO\textsubscript{3} (DSO)  & \textit{c}  & 7.903  & \textit{b}  & 5.717  & 1.30  & 2.99  \\ \hline
        LSAT  & \textit{c*2}  & 7.736  & \(\sqrt{a^2+b^2}\)  & 5.470  & -0.83  & -1.39  \\ \hline
    \end{tabular}
    \end{adjustbox}
    \caption{\textbf{Lattice parameters and resulting lattice mismatch of 001-oriented LTO on STO (110), DSO (100), and LSAT (110) substrates.} The STO (110) and LSAT (110) substrates are cubic, with epitaxy achieved by matching the LTO \textit{a} parameter to double the cubic \textit{c} parameter, while the \textit{b} parameter (aligned along the in-plane polarization, \textit{P}) matches the cubic diagonal of the length $\sqrt{a^2 + b^2}$. The DSO (100) substrate is orthorhombic, allowing the LTO \textit{a} parameter to match the DSO \textit{c} parameter, and the LTO \textit{b} parameter to align with the DSO \textit{b} parameter.}
\end{table}

\subsection*{Supporting Note S2: Effects of High Substrate Temperature on LTO Film Deposition on STO (110)}

\begin{figure}[htb!]
  \centering 
  \begin{adjustbox}{width=0.93\textwidth, center}
    \includegraphics[width=0.93\textwidth]{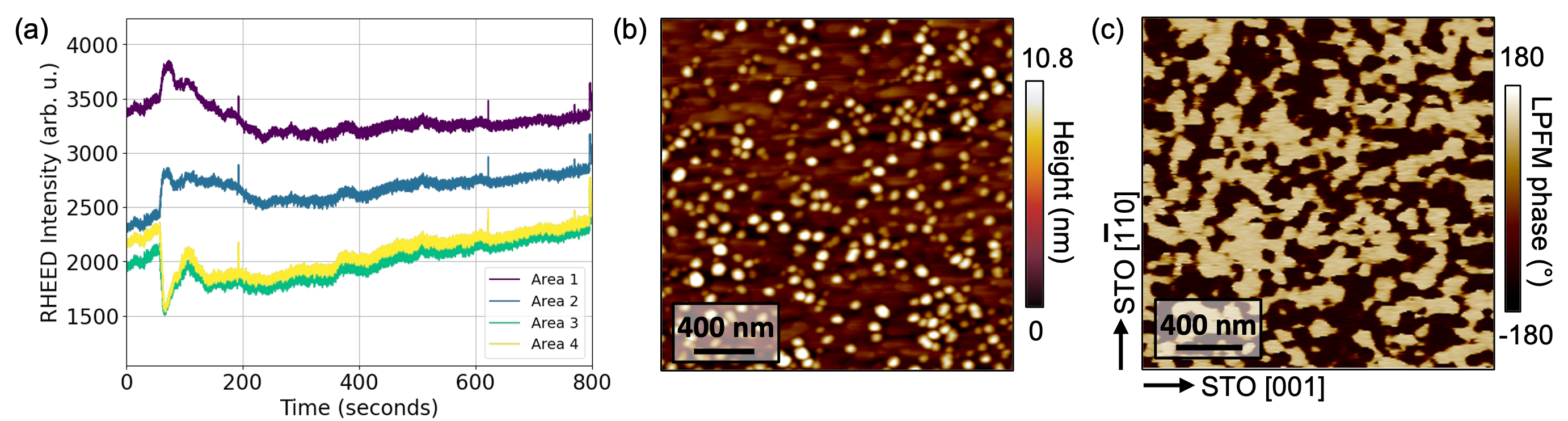}
  \end{adjustbox}
  \caption{\textbf{Impact of high substrate temperatures ($>$900\textdegree{}C) on the growth of LTO films on STO (110).} (a) Time-dependent RHEED intensity traces for different areas of an LTO film deposited at 1000\textdegree{}C. The absence of RHEED oscillations indicates a step-flow growth mode, resulting from the high kinetic energy of adsorbed species at elevated substrate temperatures. (b) Atomic force microscopy (AFM) image showing a rough surface with pronounced clusters, a consequence of splashing from target particles and step-flow growth. (c) Lateral piezoresponse force microscopy (LPFM) phase image of the LTO film on STO showing two ferroelectric in-plane domain variants, that confirm the stabilization fo the ferroelectric CG phase. These observations suggest that while higher temperatures may promote layering of the \textit{A}\textsubscript{2}\textit{B}\textsubscript{2}O\textsubscript{7} phase, the excess kinetic energy and lack of RHEED monitoring resolution make it difficult to achieve atomically smooth surfaces and precise thickness control. Therefore, a post-annealing step is preferred over direct high-temperature growth for films on this substrate to achieve optimal quality. }
\end{figure}
\newpage
\subsection*{Supporting Note S3: Vector PFM imaging of uniaxial in-plane polarization in LTO films}

\begin{figure}[htb!]
  \centering
  \includegraphics[width=0.65\textwidth]{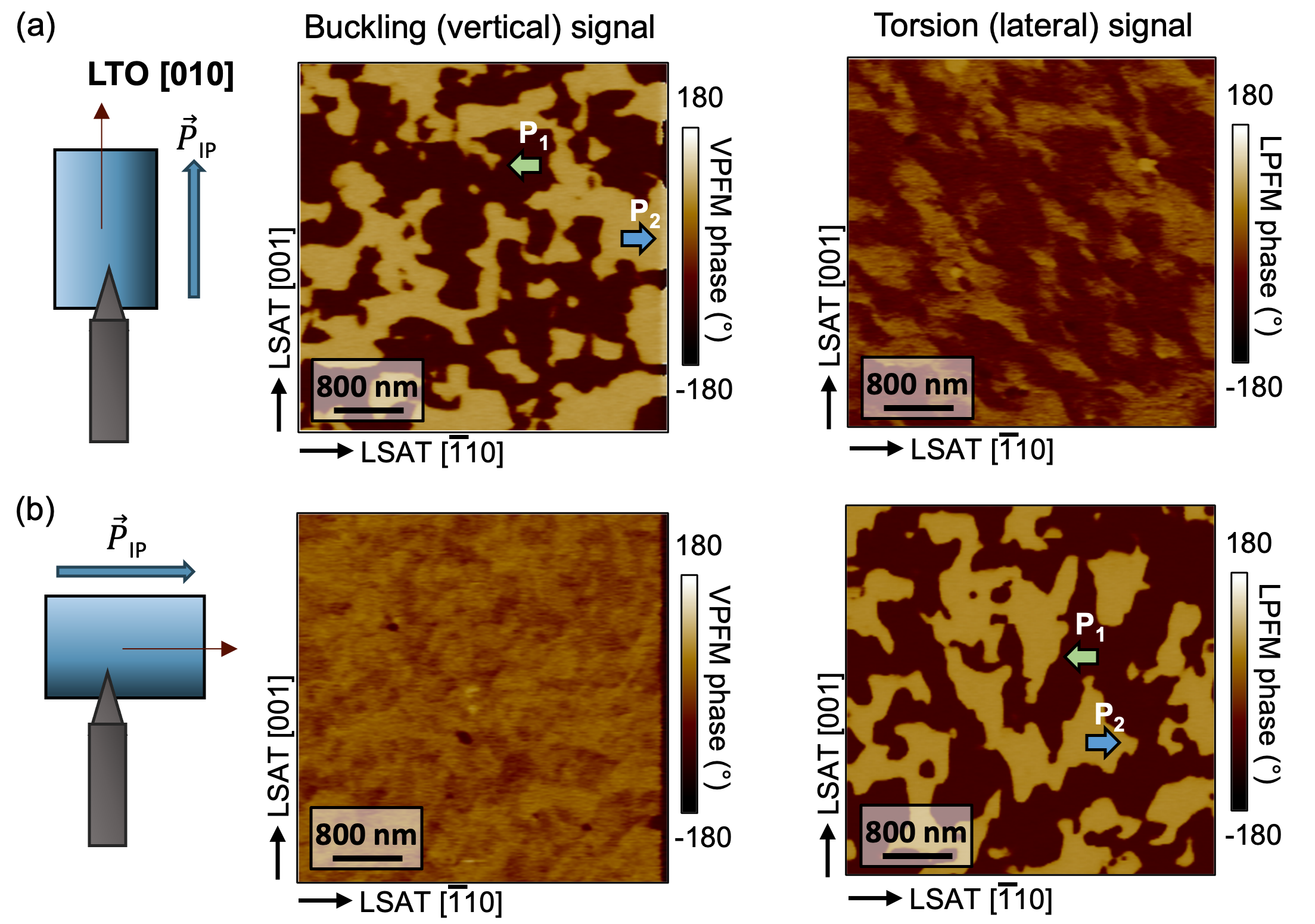}
  \caption{\textbf{Vector PFM images confirming that LTO films are uniaxially in-plane polarized.} (a) When the cantilever is parallel to the in-plane polarization (along the LTO \textit{b}-axis), the domain contrast is predominantly visible in the vertical PFM (VPFM) channel due to the cantilever's buckling mode. (b) Upon a 90\textdegree{} sample rotation, where the cantilever is now perpendicular to the in-plane polarization, the domain pattern appears in the lateral PFM (LPFM) channel due to the cantilever's torsion mode. The vertical channel now shows a homogeneous signal. The marked differences in the piezoresponse upon sample rotation validate the uniaxial in-plane polarization of the LTO films.}
\end{figure}
\newpage

\subsection*{Supporting Note S4: Phonon frequencies of the paraelectric $Cmcm$ phase of LTO}

\begin{figure}[htb!]
  \centering
  \includegraphics[width=0.75\textwidth, clip, trim=5 5 5 25]{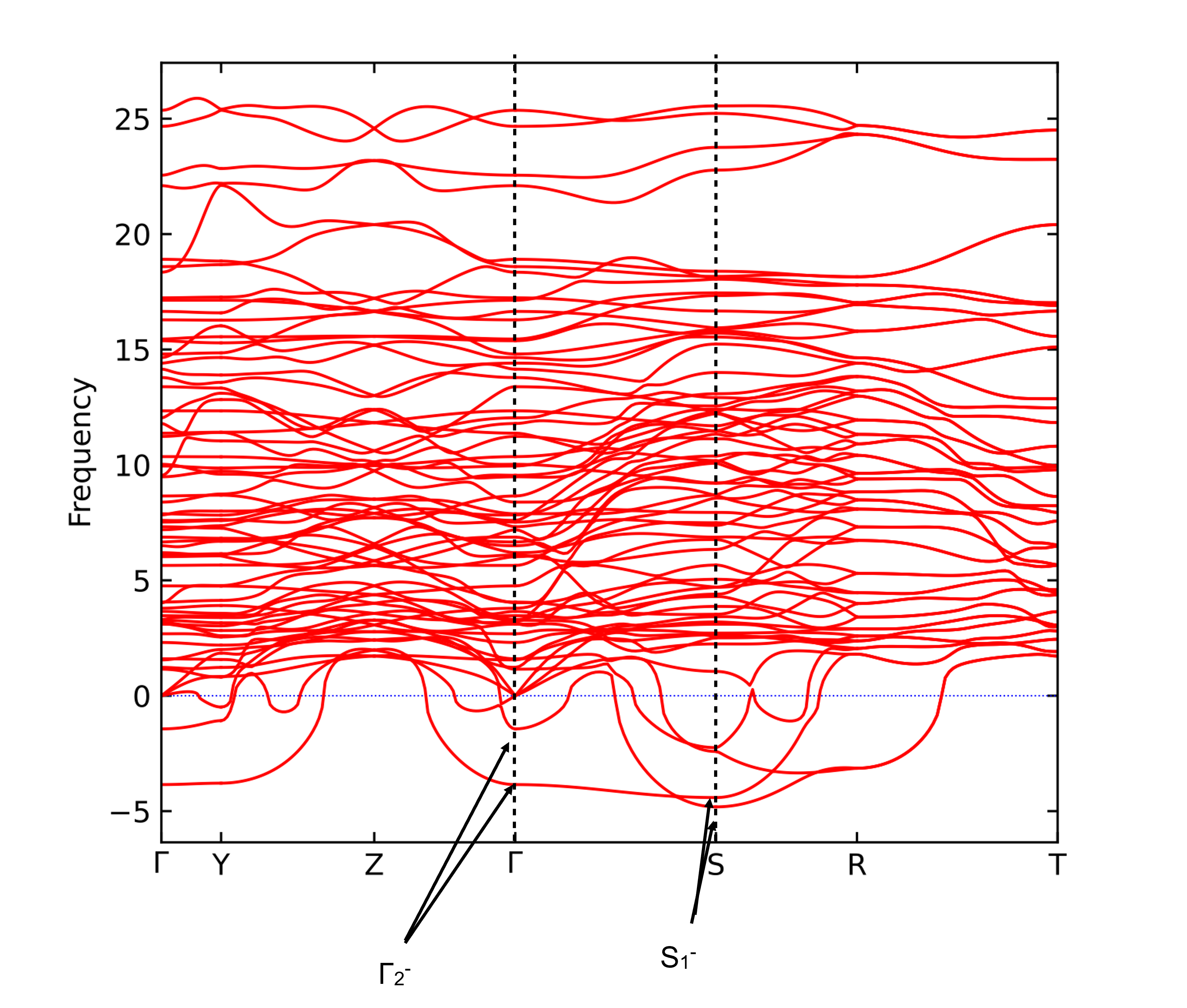}
  \caption{\textbf{Calculated phonon frequencies of LTO using DFT.} The symmetry analysis of the unstable bands shows that the two most unstable bands at the $\Gamma$ and $S$ point belong to the $\Gamma_2^-$ and $S_1^-$ irreducible representation respectively. This suggests $G_2^-$ and $S_1^-$ as the primary order parameters in the subsequent phase transitions. }
\end{figure}

\newpage

\subsection*{Supporting Note S5: Calculation of the spontaneous polarization in LTO}

\begin{figure}[htb!]
  \centering
  \includegraphics[width=0.95\textwidth, clip, trim=5 5 5 5]{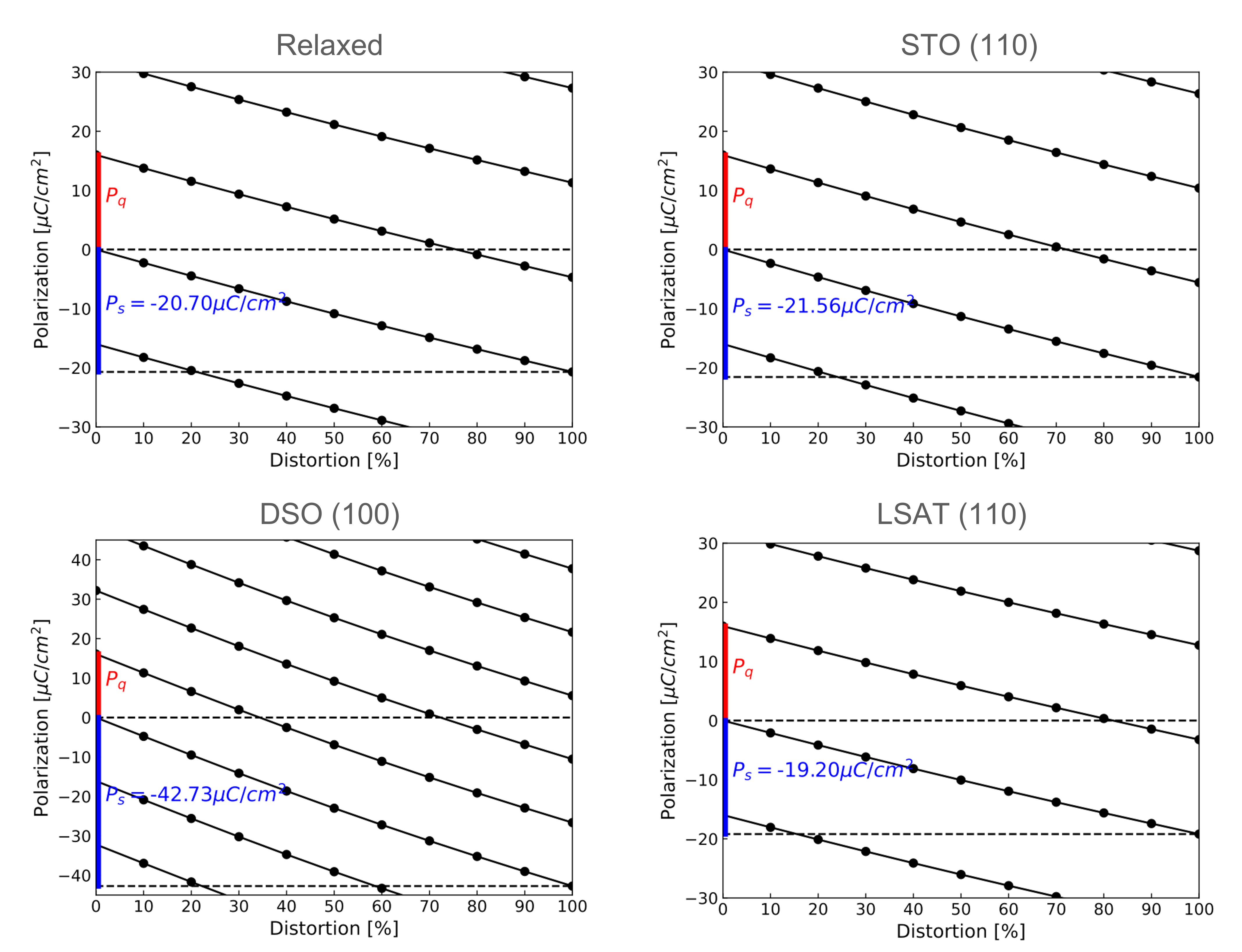}
  \caption{\textbf{Calculation of the spontaneous ferroelectric polarization} for LTO with the lattice parameters with fully relaxed lattice and internal coordinates, and with the in-plane components of the lattice constrained to the epitaxial conditions of STO, DSO and LSAT. We display the polarization lattice, the polarization quantum ($P_q$) and the resulting spontaneous polarization $(P_s)$.}
\end{figure}
\newpage

\subsection*{Supporting Note S6: Full-range X-ray diffractograms of LTO films on STO (110), DSO (100), and LSAT (110) substrates}

\begin{figure}[htb!]
  \centering
  \includegraphics[width=0.6\textwidth]{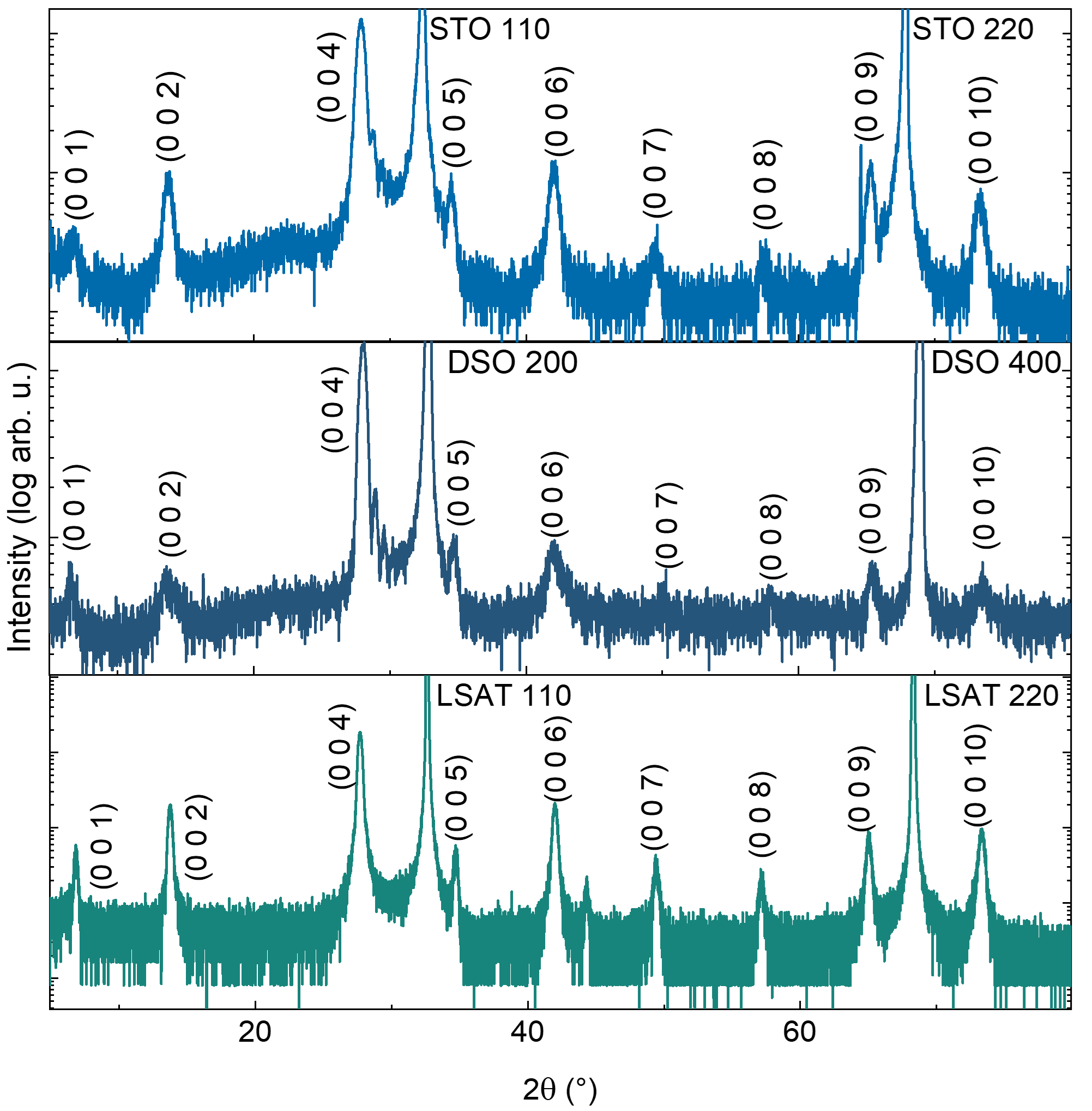}
  \caption{\textbf{Full-range X-ray diffractograms of LTO films grown on STO (110), DSO (100), and LSAT (110) substrates.} The diffractograms confirm that the LTO films are 001-oriented on all substrates, with no detectable parasitic phases.}
\end{figure}
\newpage

\pagebreak

\end{document}